\documentclass[journal]{IEEEtran} 

\usepackage{amssymb}
\usepackage{cite}
\usepackage{color,comment}
\usepackage[pdftex]{graphicx}
\usepackage{relsize}
\usepackage{amsmath}
\usepackage{authblk}
\usepackage{commath}
\usepackage{mathtools,trimclip}
\def\bba{\ensuremath{\mathbf a}}
\def\bbb{\ensuremath{\mathbf b}}

\def\bbs{\ensuremath{\mathbf s}}
\def\bbS{\ensuremath{\mathbf S}}

\def\bbP{\ensuremath{\mathbf P}}

\def\bbr{\ensuremath{\mathbf r}}

\def\bbS{\ensuremath{\mathbf S}}

\def\bbH{\ensuremath{\mathbf H}}

\def\bbW{\ensuremath{\mathbf W}}

\def\bbY{\ensuremath{\mathbf Y}}

\def\bbZ{\ensuremath{\mathbf Z}}

\begin{document}
\title{\LARGE GR-WiFi: A GNU Radio based WiFi Platform with Single-User and Multi-User MIMO Capability}

\author{Natong~Lin$^1$, Zelin~Yun$^1$, 
        Shengli~Zhou$^2$,~\IEEEmembership{Fellow,~IEEE,}
        and~Song~Han$^1$,~\IEEEmembership{Member,~IEEE}\\

{$^1$School of Computing, University of Connecticut}\\
{$^2$Department of Electrical and Computer Engineering, University of Connecticut } \\
{Email: \{natong.lin, zelin.yun, shengli.zhou, song.han\}@uconn.edu}
}

\maketitle

\bigskip
\begin{abstract}

Since its first release, WiFi has been highly successful in providing wireless local area networks. The ever-evolving IEEE 802.11 standards continue to add new features to keep up with the trend of increasing numbers of mobile devices and the growth of Internet of Things (IoT) applications. Unfortunately, the lack of open-source IEEE 802.11 testbeds in the community limits the development and performance evaluation of those new features. 
Motivated by an existing popular open-source 
software-defined radio (SDR) package for single-user single-stream transmission based on the IEEE 802.11/a/g/p standard, in this paper we present GR-WiFi, an open-source package for single-user and multi-user multi-input multi-output (MIMO) transmissions based on 802.11n and 802.11ac standards. 
The distinct features of GR-WiFi include the support of parallel data streams to single or multiple users, and the compatible preamble processing to allow the co-existence of conventional, high-throughput (HT) and very-high-throughput (VHT) traffics. The performance of GR-WiFi is evaluated through both extensive simulation and real-world experiments.

\end{abstract}

\section{Introduction}

The rapid advancement of IoT technologies has driven significant research efforts toward developing flexible and scalable communication frameworks. IoT applications leverage a variety of wireless communication standards, with IEEE 802.15.4 and IEEE 802.11 being the representative ones~\cite{8024171,sisinni2018industrial}. 
Significant progress has been made in the development of software-defined radio (SDR) implementations of IEEE 802.15.4. Notable examples include
GNU Radio packages on the SDR platforms
based on noncoherent receivers 
\cite{bloessl2013gnu} 
and an coherent receiver \cite{9348953}. While IEEE 802.15.4 remains crucial for low-power and short-range IoT applications, the increasing demand for higher data rates and enhanced throughput has shifted research attention toward IEEE 802.11.

The IEEE 802.11 standard was first released in 1997 designed for wireless local area network (WLAN) usage as part of the IEEE 802 family. With advancements in integrated circuits, IEEE 802.11 has become the foundation for WiFi products and one of the most widely adopted wireless communication standards globally. Following the first widely used version IEEE 802.11b (WiFi 1) in 1999, the 802.11 working group (WG) continued to evolve and release the versions of IEEE 802.11a (WiFi 2) and IEEE 802.11g~(WiFi 3)  \cite{gast2005802} supporting orthogonal frequency division multiplexing (OFDM). In IEEE 802.11n~(WiFi 4) \cite{9502043}, the single-user multiple input multiple output (SU-MIMO) is supported with transmissions of parallel data streams. In IEEE 802.11ac~(WiFi 5) \cite{9502043}, multi-user MIMO (MU-MIMO) is added which relies on 
channel feedback and linear precoded transmissions. 
In IEEE 802.11ax~(WiFi 6) released in 2021, 
orthogonal frequency division multiple access (OFDMA) is supported 
as an option for multiuser transmissions.

While WiFi hardware vendors continually release products aligned with the latest IEEE 802.11 standards, commercial off-the-shelf (COTS) hardware is typically not open-source, and the fixed functions of application-specific integrated circuits (ASICs) limit their use in research and performance evaluation. Although the open-source community provides drivers for some COTS hardware (e.g., AR9280 and RTL8812), these drivers are limited in functionality, lack updates, and may not remain compatible with evolving OS kernels and 802.11 standards. Project like Nexmon \cite{SCHULZ2018269}, offers an alternative by enabling custom firmware modifications and enhancements at a lower level. However, its support is limited to specific Broadcom/Cypress chipsets and often requires advanced technical expertise for effective implementation.

 \begin{figure*}[t]
 \centering
 \includegraphics[width=1\linewidth]{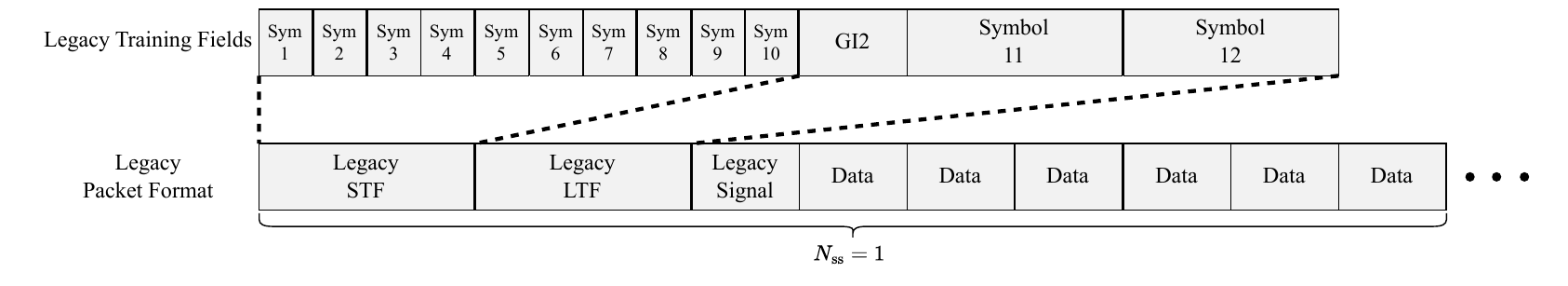}
\vspace{-0.3in}
\caption{IEEE 802.11a/g with one spatial 
stream, which could be transmitted by
multiple antennas via cyclic shifts}
\label{fig:80211a_g}
 \end{figure*}

\begin{figure*}[t]
 \includegraphics[width=1\linewidth]{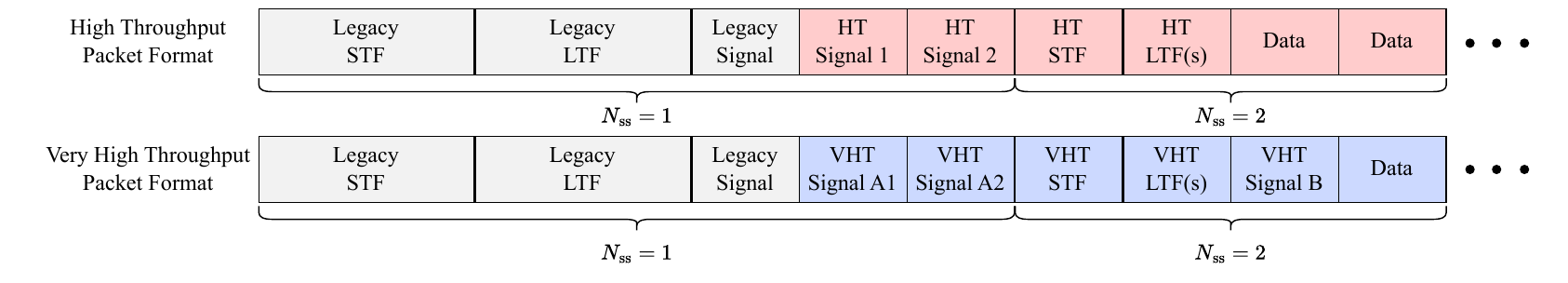}
 \vspace{-0.3in}
 \caption{IEEE 802.11n and 802.11ac packet formats with parallel data streams. 
 }
 \label{fig:phyformats}
 \end{figure*}

Among various existing work of open-source implementations
of IEEE 802.11 standards,  the work in this paper is primarily motivated by a widely-used open-source 
project, WIME \cite{bloessl2018performance}, which  
provides a GNU Radio based package for SDR platforms.
However, the open-source package in the WIME project only supports single data stream transmission defined 
in IEEE 802.11a/g/p. 
To overcome this limitation, we have 
developed an open-source GNU Radio package, 
GR-WiFi~\cite{grwifi},
which 
supports the 2\texttimes2 
single user MIMO transmission introduced in IEEE 802.11n and the 2\texttimes2
multi-user MIMO introduced in IEEE 802.11ac.
Thanks to the unique preamble design for backward compatibility, the GR-WiFi package supports the the co-existence of Legacy OFDM (Legacy), high-throughput (HT), and very-high-throughput (VHT) formats seamlessly. 
These advanced MIMO and physical layer capabilities enhance the efficiency of wireless communication and provide researchers and developers a 
convenient and flexible
open-source environment to evaluate performance 
and experiment with next-generation network technologies. 

To validate the functionality and performance of GR-WiFi, we conduct both simulation and testbed evaluations. In simulations, we assess system performance by introducing interferences, including additive white Gaussian noise (AWGN), carrier frequency offset (CFO) and TGac channel models, to test robustness under controlled conditions. For testbed evaluations, we perform experiments to evaluate the effectiveness of GR-WiFi under practical operating conditions. 

The remainder of this paper is organized as follows. Section~\ref{sec:realated} provides a summary of the related work. Section~\ref{sec:system} presents the HT and VHT packet formats. Section~\ref{sec:receiver} describes key receiver modules. Section~\ref{sec:implement} provides the details of the system implementation in GNU Radio. Section~\ref{sec:performance} contains the simulation and testbed setup and demonstrates the performance of the implemented system. Section~\ref{sec:conclusion} concludes the paper with a discussion of 
ongoing and future work.

\textit{Notation:} Lower-case letters stand for time domain and upper-case letters stand for frequency domain symbols in OFDM transmissions. Boldface letters stand for vectors and matrices. Symbol $|\cdot|$ stands for the absolute value, $\Vert \cdot \Vert$ for the 2-norm of a vector, and $(\cdot)^H$ for the Hermitian transpose of a vector or matrix. 
For two row vectors $\bba$ and $\bbb$, the correlation coefficient is defined as
\begin{equation}
<\bba,\bbb> \; = \; \frac{|\bba \bbb^H|}{ \Vert\bba\Vert \cdot \Vert\bbb\Vert }.
\end{equation}
\section{Related Work}\label{sec:realated}
As WiFi standards continue to evolve, maintaining scalable and adaptable testbeds is essential for evaluating new designs and implementations. The Openwifi project \cite{openwifi} offers a software-defined IEEE 802.11a/g/n implementation on FPGA, providing a Linux MAC80211-compatible open-source WiFi communication stack. The BladeRF-wiphy project \cite{nuandBladeRFwiphyNuand} provides an FPGA-based open-source IEEE 802.11a/g SDR modem that primarily focuses on the physical layer (PHY) implementation, facilitating modulation and demodulation of 802.11 packets.  The Wireless Measurement and Experimentation (WIME) project \cite{WIME} complements these efforts by offering comprehensive SDR-based framework supporting both IEEE 802.11 and IEEE 802.15.4 protocols. WIME integrates USRP devices with GNU Radio, creating a scalable testbed for wireless protocol development and performance evaluation. These platforms provide flexible open-source tools that enable researchers to explore single-user system and pursue advanced projects, such as RT-WiFi \cite{9804669}, which extends these efforts by introducing a real-time WiFi framework designed for industrial applications, achieving deterministic communication performance and high-speed real-time data transmission. Although these above projects provide flexible open-source platforms and tools that empower researchers to explore single-user system configurations, they fall short in addressing more advanced multi-user scenarios. Specifically, the development of a multi-user MIMO (MU-MIMO) testbed lacks comprehensive open-source solutions. 

Despite this limitation in open-source offerings, there have been multiple related works on MU-MIMO implementation in other contexts. For instance, an MU-MIMO testbed was developed in \cite{4022536} for two users in an indoor environment, supporting packet transmission in legacy format, but the channel state information (CSI) used for determining the transmit weights is transferred through a wired connection. 
The FPGA-based testbed in \cite{4489270} enables real-time CSI feedback and closed-loop beamforming, however, it follows the signaling format with 3G LTE standard rather than WiFi. 
Multi-user beamforming experiments were carried out in \cite{2010mobicom}, using one 
narrowband channel corresponding 
to one OFDM subcarrier. 
MU-MIMO experiments can be carried out
in the ORBIT (Open-Access Research Testbed for Next-Generation Wireless Networks) testbed as
outlined in \cite{8116431}. 
However, offline CSI 
processing was deployed 
where the beamforming matrix is computed in Matlab after channel sounding. 
A learning-based channel 
feedback framework was proposed
and experimentally evaluated 
in \cite{9259366} for MU-MIMO 
in WLANs. There are also testbeds built on top of COTS devices such as \cite{10.1145/2973750.2973758}, which utilizes commodity 802.11ac APs to develop a MU-MIMO testbed that investigates a novel user selection algorithm called MUSE, addressing practical challenges in real-world deployments. 
However, none of the works in \cite{4022536,8116431,4489270,9259366,2010mobicom,10.1145/2973750.2973758} is available as open-source packages.

In MU-MIMO systems, channel estimation is critical since it helps the transmitter decide spatial multiplexing and rate selection. A distinctive feature of our GR-WiFi platform is its real-time MU-MIMO capability, establishing a communication link between AP and stations without needing a wired connection for CSI exchange. This enables more realistic, flexible testing, 
paving the way for new advancements in next-generation Wi-Fi research. Another distinctive feature of the GR-WiFi package is the support of co-existence of legacy, HT, and VHT terminals seamlessly through the preamble designed for mixed-formats in IEEE 802.11n/ac.
\section{HT and VHT Packet Formats}\label{sec:system}

Fig.~\ref{fig:phyformats}
shows the packet formats for 
HT transmissions in
IEEE 802.11n and VHT transmissions in IEEE 802.11ac,
which highlights the connection and 
the distinctions of the conventional transmission
in IEEE 802.11a/g in Fig.~\ref{fig:80211a_g}.
While the legacy OFDM transmission has only one spatial data stream ($N_{\rm ss}=1$),
the HT and VHT packets consists of a 
preamble of one spatial data stream
and the data transmission of two spatial streams ($N_{\rm ss}=2$). Through the use of cyclic shift diversity (CSD), even the preamble is transmitted through  multiple antennas. The preamble is designed such that 
legacy, HT and VHT traffics can be seamlessly supported. Due to the hardware limitation, we have only implemented the 20 MHz bandwidth and up to 2\texttimes2 MIMO in this paper. For brevity, we next provide a description of the packet formats with two antennas at the transmitter.

\subsection{Preamble via Cyclic Shift Diversity }

The preamble portion of the HT and VHT packets has only one spatial data stream. To map the single data stream to the transmitter with multiple antenna elements, the CSD concept 
is adopted, which prevents unintended beamforming when the same signal is transmitted through a different antenna.

Consider a generic (say, the $n$th) OFDM data block in the preamble at 
$t \in [nT, n(T +T_{\rm GI})]$: 
\begin{equation}
    \begin{aligned}
    x(t) & = 
    \sum_{k}^{} S[n;k] e^{j2\pi k \Delta_{\rm F} (t-nT_{\rm bl}-T_{\rm GI})}, 
    \end{aligned}
\end{equation}
where $k$ is the subcarrier index, $S[n;k]$ is the information symbol on subcarrier $k$, $\Delta_{\rm F}$ is the subcarrier frequency spacing, $T$ symbol duration, $T_{\rm GI}$ is the guard interval (GI) for each symbol, and the block duration $T_{\rm bl}=T+T_{\rm GI}$. The transmitted waveform is cyclically delayed by $T_i^{\rm CS}$ on different antennas:
\begin{equation}
x_{i}(t)
=\begin{cases} 
    x(t-T_i^{\rm CS}), 
    & nT_{\rm bl} \le t < n(T_{\rm bl}+T_i^{\rm CS}) \\
    x(t-T_i^{\rm CS}-T), &
    n(T_{\rm bl}+T_i^{\rm CS}) \le t \le (n+1)T_{\rm bl}
\end{cases}
\label{eq:mimotxmodel}
\end{equation} 
where $T_i^{\rm CS}\le 0$ is assumed. 
For two antennas, we 
have $T_1^{\rm CS}=0$ ns and $T_2^{\rm CS} = -200$ ns. 
The preamble has the following parts.

\begin{figure}[t]
    \centering
    \includegraphics[width=0.9\columnwidth]{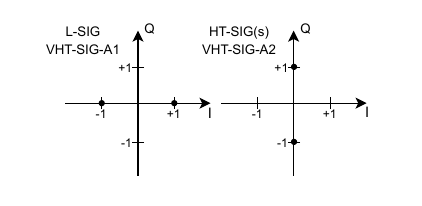}
    \caption{BPSK constellation (left) and QBPSK constellation (right) 
    }
    \label{fig:sigconstellation}
\end{figure}

\begin{itemize}
    \item Legacy Preamble.
    The legacy preamble part for HT and VHT are identical as in legacy OFDM transmission,
    which includes legacy short training field (L-STF), legacy long training field (L-LTF), and legacy signal (L-SIG). L-STF consists of ten repeated symbols spanning 8 $\mu s$. L-LTF includes two repeated symbols and guard interval GI2 of 8 $\mu s$. The subcarrier sequences for training symbols are given by the standard. L-SIG consists of 24-bit of the rate and length information. Binary convolutional coding (BCC) and interleaving are applied to the bit stream, and 
    binary-phase-shift-keying (BPSK) 
    constellation is used for bit-to-symbol mapping. For HT/VHT transmission, the rate is set to lowest, and the length is computed to cover the time for transmitting the following HT/VHT portion. 

    \item After the legacy preamble, the HT transmission contains two HT-SIG symbols: HT-SIG1 and HT-SIG2. HT-SIG1 indicates  the modulation and coding scheme (MCS) value and the payload length in octets. In our $2 \times 2$ setting, MCS has a value of 8 to 15, representing two spatial streams using equal modulation (EQM). HT-SIG2 has the cyclic redundancy check (CRC) to verify the integrity of the HT-SIG and detect errors. Instead of BPSK constellation, both HT-SIG signals map the encoded and interleaved data bits into quadrature-binary-phase-shift-keying (QBPSK) modulation, as shown in Fig.~\ref{fig:sigconstellation}. 

    \item After the legacy preamble, the VHT transmission contains two VHT-SIG symbols: VHT-SIG-A1 and VHT-SIG-A2. VHT-SIG-A1 decides whether the packet is in MU-MIMO or SU-MIMO mode. VHT-SIG-A2 indicates a coding scheme for each user and also has CRC to verify the integrity of the VHT-SIG and detect errors. In our setting, only BCC is supported. Different from HT signaling parts, VHT-SIG-A1 is mapped to BPSK
    modulation, and VHT-SIG-A2 
    is mapped to QBPSK, as shown in Fig.~\ref{fig:sigconstellation}.

\end{itemize}

\subsection{SU-MIMO Transmission}
Following the single-stream preamble, the HT transmission afterwards has parallel data streams. HT-STF, HT-LTF, are inserted in front of the data transmission for synchronization and channel re-estimation. 
Denote $S_i[n;k]$ as the information symbol on the $k$th subcarrier
of the $n$th symbol on antenna $i$.
For the $n$th OFDM block with $t\in [nT_{\rm bl},  (n+1)T_{\rm bl}]$,  
the OFDM block is written as 
\begin{equation}
     \begin{aligned}
     x_i(t) & = 
     \sum_{k}^{} S_{i}[n;k] e^{j2\pi k \Delta_{\rm F} (t-T_{\rm GI}-T_{i}^{\rm CS})} 
     \end{aligned}
 \end{equation}
where CSD is applied with  
$T_1^{\rm CS}=0$ ns and 
$T_2^{\rm CS} = -400$ ns.

\begin{figure}[t]
    \includegraphics[width=1.0\columnwidth]{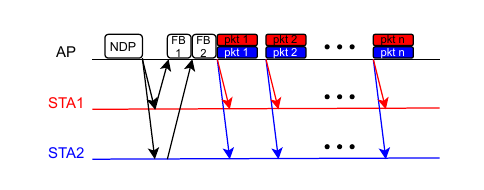}
    \caption{Diagram of the MU-MIMO transmission procedure.
    }
    \label{fig:diagram}
\end{figure}

\subsection{MU-MIMO Transmission}
 
MU-MIMO transmission in IEEE 802.11ac enables data transmission to multiple users simultaneously by leveraging spatial divsion multiple access (SDMA). 
The process begins with channel sounding, where the AP sends a Null Data Packet (NDP) to allow users to measure their channel conditions and provide feedback. Based on this feedback, the AP calculates beamforming weights to focus each data stream toward its intended user while minimizing interference to others. By applying these weights during transmission, the AP can deliver multiple spatially separated data streams at the same time. 

In our two-user testbed implementation, the transmission procedure is described in Fig.~\ref{fig:diagram}. AP first sends NDP packets to sound the channel. Black arrow indicates omi-directional transmission. Station 1 (STA1) responds by sending the channel feedback (FB) back to the AP immediately, while Station 2 (STA2) waits for a pre-specified time before sending FB back. Next, the AP computes and applies precoding into two spatial streams. Let $S^{i}[n;k]$ be the intended symbol to station $i$.
On each subcarrier $k$, two symbol streams are generated as 
\begin{equation}
\begin{bmatrix}
    \tilde S_1[n;k] \\ 
    \tilde S_2[n;k] 
\end{bmatrix}
=
   \begin{bmatrix}
        Q_{11}[k] & Q_{12}[k] \\
        Q_{21}[k] & Q_{22}[k] 
    \end{bmatrix} 
\begin{bmatrix}
     S^{(1)}[n;k] \\ 
     S^{(2)}[n;k] 
\end{bmatrix}.
\end{equation}
The precoding operation is applied on 
VHT-STF, VHT-LTF, VHT-Sig-B, and data packets.
For the $n$th OFDM block with $t \in [nT_{\rm bl}, (n+1)T_{\rm bl}] $, 
the following OFDM symbols are generated 
for antenna $i$: 
\begin{equation}
     \begin{aligned}
     x_i(t) & = 
     \sum_{k}^{} \tilde S_{i}[n;k] e^{j2\pi k \Delta_{\rm F} (t-nT_{\rm bl}- T_{\rm GI}-T_{i}^{\rm CS})},
     \end{aligned}
\end{equation}
where the same cyclic shifts are applied as in HT transmission. The precoded symbol streams are sent out from the AP while two stations at different locations decode their own symbols separately; in Fig.~\ref{fig:diagram}, the red arrow indicates spatial stream for STA1 and blue for STA2. 
\section{Key Receiver Modules}\label{sec:receiver}

\begin{figure}[t]
    \centering
    \includegraphics[width=1\columnwidth]{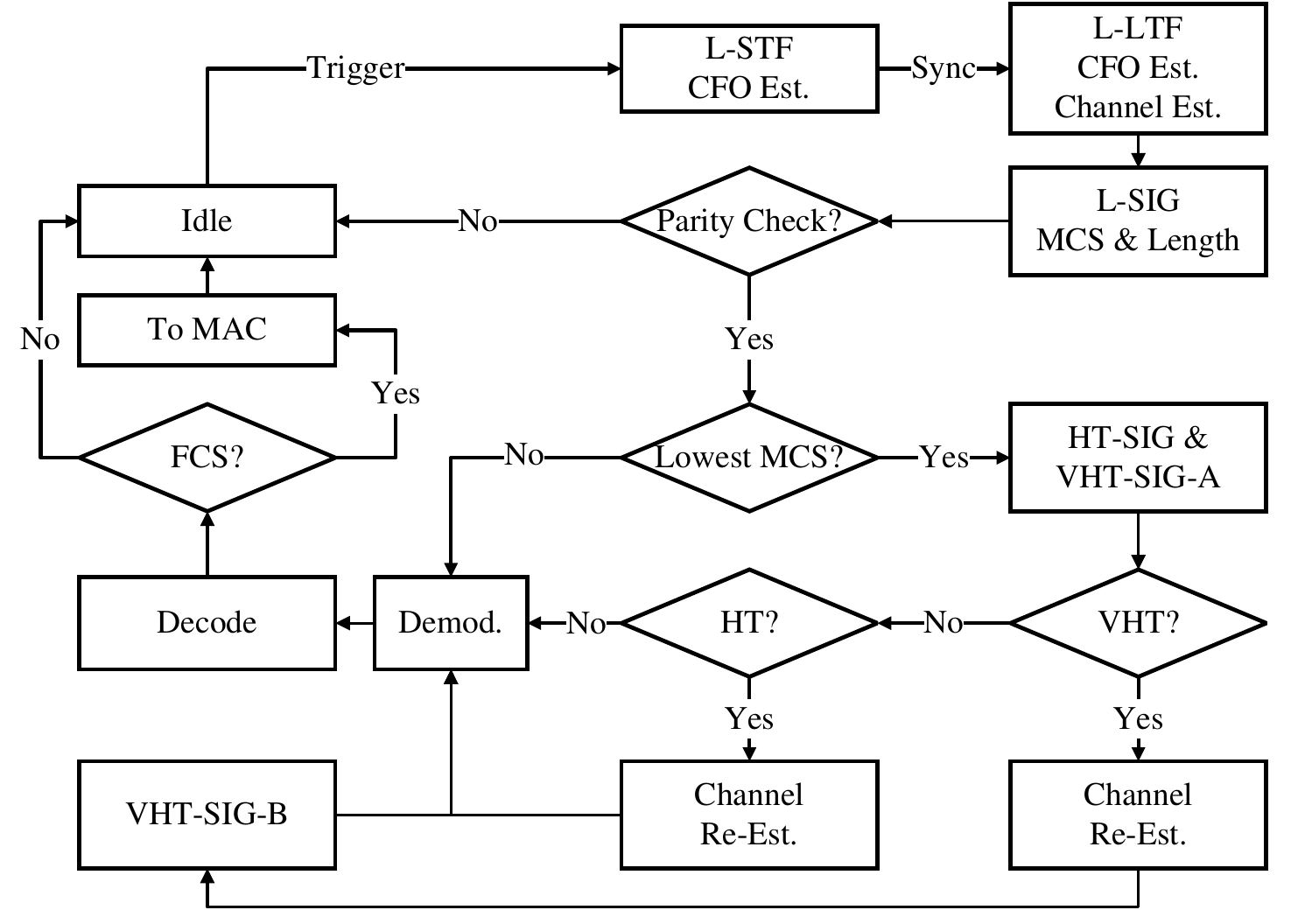}
    \caption{State machine of the proposed receiver for all packet formats.}
    \label{fig:statemachine}
\end{figure}

Fig.~\ref{fig:statemachine} depicts the 
state machine used by the GR-WiFi system during
the packet reception, which will first determine
the packet format before processing the data blocks. For Legacy packets, the receiver uses the L-STF and L-LTF fields for synchronization and channel estimation. The L-SIG field provides the MCS and payload length, enabling the receiver to decode the data payload.

For HT packets, the receiver initially treats the packet as Legacy and checks for HT compatibility by processing two additional data symbols following the L-SIG. If HT packet is detected, the receiver redoes the automatic gain control (AGC) and re-estimates the channel to account for potential differences in symbol length, bandwidth, and MIMO settings. The additional HT-STF and HT-LTF fields facilitate this process, allowing precise channel estimation for MIMO transmissions.

Similarly, for VHT packets, the state machine identifies the packet format by decoding VHT-SIG-A1 and
VHT-SIG-A2. 
The VHT-LTF training fields are then used to re-estimate the channel. The receiver further processes VHT-SIG-B to extract individual MCS and payload length for multi-user MIMO. This ensures that each user decodes only its assigned spatial streams, enabling efficient and accurate packet reception in multi-user scenarios. 

This section presents the key receiver modules, which address the challenges specific to MIMO reception in synchronization, channel estimation, and interference suppression. We begin by detailing the SU-MIMO reception process, followed by a discussion of MU-MIMO reception. 

\subsection{Processing the Single-Stream Preamble}

The receiver detects and processes the preamble based on the signal 
from one receive antenna. Hence, we omit the index on the receive antenna in this subsection.
With $x_i(t)$ in \eqref{eq:mimotxmodel}, the received signal is:

\begin{equation}
\label{eq:mimomodel}
    r(t) = \sum_{i=1}^{N_{\rm TX}} x_{i}(t) \star h_{i}(t) + w(t). 
\end{equation}
Due to the cyclic shift operation, one can define 
an equivalent channel as
\begin{equation}
h_{\rm equ}(t) = 
\sum_{i=1}^{N_{\rm TX}} h_{i}(t-T_{i}^{\rm CS}), 
\end{equation}
and re-write \eqref{eq:mimomodel} as
\begin{equation}
    r(t) = x(t) \star h_{\rm equ}(t) + w(t).
\end{equation}
Hence, for the preamble part, the receiver deals 
with \textit{an equivalent channel that consists of contributions from individual multipath channels corresponding to multiple transmit-antennas.}

A sampling rate of $20$ MHz is used, which leads to discrete samples of

\begin{equation}
    r[n] = r(t)|_{t= nT/K}.
\end{equation}
For each OFDM symbol of duration $T_{\rm bl}=4.0\ \mu s$, there are 80 discrete samples. 

\begin{figure}[t]
\centering
    \includegraphics[width=1\columnwidth]{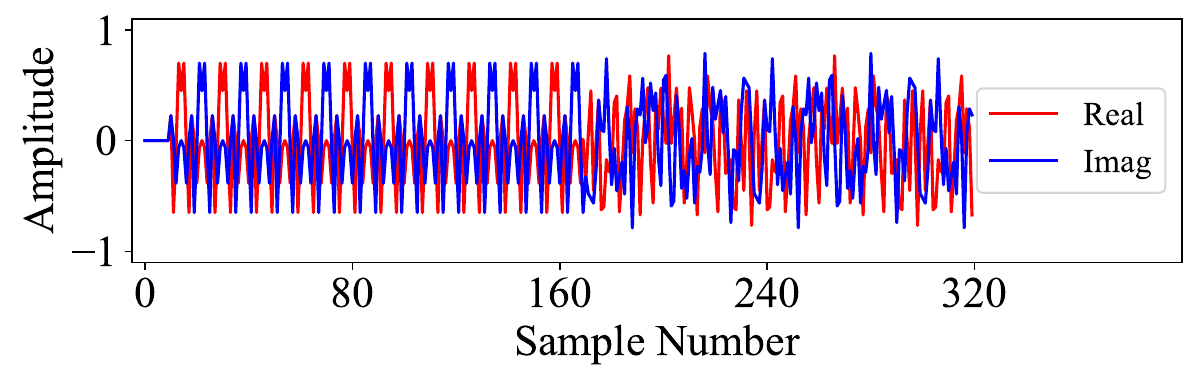}
    \caption{Legacy STF of 160 sampls and legacy LTF of 160 samples.}
    \label{fig:legacytrain}
\end{figure}

\begin{figure}[t]
\centering
    \includegraphics[width=1\columnwidth]{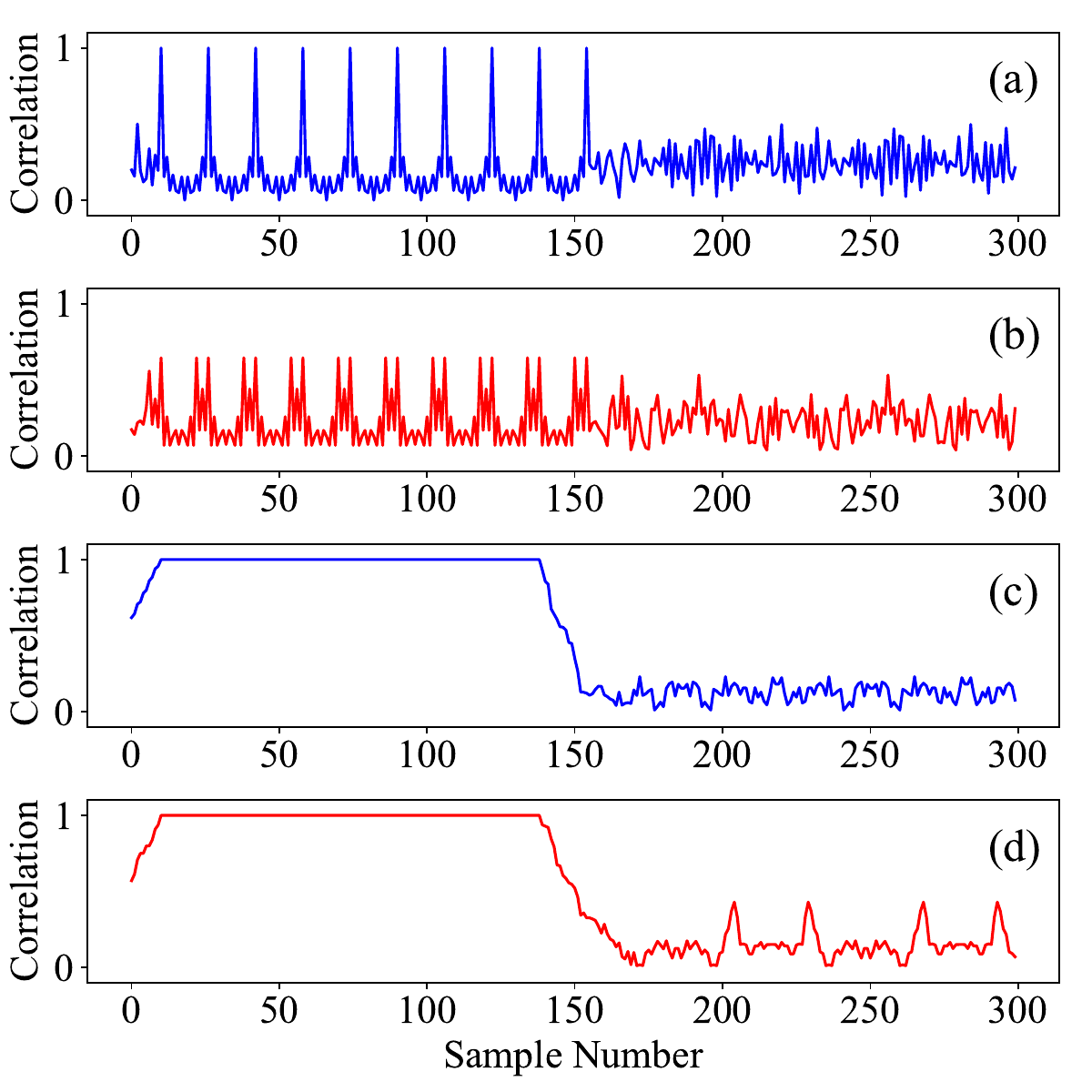}
    \caption{(a) Cross-correlation of a SISO packet Legacy training field and STF symbol; one antenna at the transmitter. (b) Cross-correlation of a 2\texttimes2 MIMO packet Legacy training field and STF symbol; two antennas at the transmitter (c) Auto-correlation of a SISO packet Legacy training field; one antenna at the transmitter. (d) Auto-correlation of a 2\texttimes2 MIMO packet Legacy training field; two antennas at the transmitter.}
    \label{fig:legacytrigger}
\end{figure}

\medskip
\subsubsection{Packet Detection via L-STF}
Fig.~\ref{fig:legacytrain} shows the legacy STF and LTF parts of the preamble.
Legacy LTF consists of 10 identical short symbols, each having $N_{\rm STF}=16$ samples. Define $\bbs_{\rm STF}$ as
the row vector containing the 16 samples of the short symbol.
Define a row vector of length $N_{\rm STF}$ at each index $n$:
\begin{equation}
    \bbr[n] = 
    \big[r[n], \ldots, r[n+N_{\rm STF}-1]\big].
\end{equation}

For packet detection, the receiver can either rely on either a cross-correlation or an auto-correlation method.
From cross correlation, the receiver computes
 
\begin{equation}
    \rho_{\rm CC}[n] = <\bbr[n], \bbs_{\rm STF}> ,
\end{equation}
utilizing the known template.
For auto correlation, the receiver computes  \begin{equation}
    \rho_{\rm AC}[n] = <\bbr[n], \bbr[n + N_{\rm STF}]> ,
\end{equation}
utilizing the repetition structure. It turns out that 
whether one should favor cross-correlation or auto correlation 
depends on the number of antennas at the transmitter employing the cyclic shifts.

Fig.~\ref{fig:legacytrigger}(a) and (b) show $\rho_{\rm CC}[n]$
for an SISO packet with one antenna at the transmitter and a MIMO packet with two antennas at the transmitter, respectively,
where no noise or interference or carrier frequency offset (CFO) is assumed. 
It can be seen that the cross correlation output in Fig.~\ref{fig:legacytrigger}(b) shows more peaks but with lower heights
than that in 
Fig.~\ref{fig:legacytrigger}(a).
Setting a threshold for finding large correlation peaks becomes more challenging for a MIMO packet with CSD.
On the other hand, 
Fig.~\ref{fig:legacytrigger}(c) and (d) show $\rho_{\rm AC}[n]$ for a SISO and a MIMO packet, respectively. With a given threshold, the receiver counts the continuous plateau length to declare packet detection. Comparison of Fig.~\ref{fig:legacytrigger}(c) and (d) suggests that the auto correlation method provides more consistent performance for both SISO and MIMO packets.
We hence adopt the  autocorrelation method for performance testings. 

\begin{figure}[t]
\centering
    \includegraphics[width=1\columnwidth]{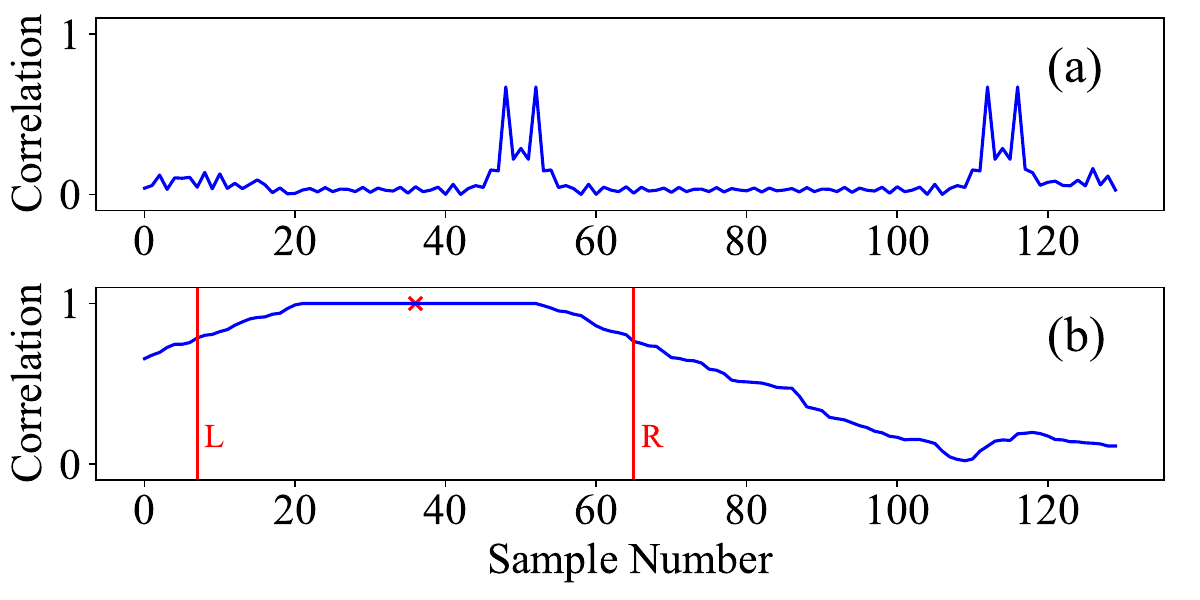}
    \caption{(a) Cross-correlation of a 2\texttimes2 MIMO packet Legacy LTF field and LTF symbol. (b) Auto-correlation of a 2\texttimes2 MIMO packet LTF field.}
    \label{fig:legacysync}
\end{figure}

Assume that the index $\hat n_1$ marks a coarse start time of L-STF of a detected packet, a coarse estimate of 
CFO can be calculated as 

\begin{equation}
\hat\epsilon_{1} = \frac{ 
\text{phase}
\left\{\sum_{l=0}^{L-1} 
(r[\hat n_1+l])^*
r[\hat n_1 + l+ N_{\rm STF}]\right\}}
{N_{\rm STF}} 
\end{equation} 
which involves 
a segment of $L+N_{\rm STF}$ samples from STF. It was set
$L= 3N_{\rm STF}$,
but the value could be adjusted by the users. 
Coarse CFO compensation is carried out to generate 
data samples 
\begin{equation}
\tilde r[n]= 
e^{-j n \hat \epsilon_1} r[n],
\end{equation}
for down-steam processing 
on legacy LTF.

\medskip
\subsubsection{Processing L-LTF} 
The L-LTF consists of two identical symbols, each of $N_{\rm LTF}=64$ samples, prepended by 
$32$ samples in the guard interval.
Define $\bbs_{\rm LTF}$ as the row vector containing these 64 samples.
Define a row vector of length $N_{\rm LTF}$ at each index $n$
\begin{equation}
    \tilde \bbr[n] = \big[\tilde r[n], \ldots, \tilde r[n+N_{\rm LTF}-1]\big].
\end{equation}
Once again, the receiver might adopt the cross correlation or auto correlation the method for fine synchronization by computing 
\begin{align}
    \tilde\rho_{\rm CC}[n] &= <\tilde \bbr[n], \bbs_{\rm LTF}>, \\ 
    \tilde\rho_{\rm AC}[n] &= <\tilde\bbr[n], 
    \tilde\bbr[n + N_{\rm LTF}]> .
\end{align}

Fig.~\ref{fig:legacysync}(a) and
(b) show $\tilde \rho_{\rm CC}[n]$
and $\tilde \rho_{\rm AC}[n]$ for a MIMO packet with two antennas at the transmitter, respectively.
Similar to the discussion for L-STF, 
cross correlation output has doubled the number of peaks with smaller amplitudes 
when the cyclic shift is applied.
On the other hand, 
the auto-correlation result outputs a desriable plateau. The fine-time point is found by identifying the maximum auto-correlation within the LTF range and then determining the shoulder index where the shoulders are the auto-correlations mostly close to the 80\% of the maximum value on both left and right sides of the maximum value as shown in Fig.~\ref{fig:legacysync}(b). 
Based on the format in Fig.~\ref{fig:phyformats}, 
the auto correlation is expected to reach the maximum value starting from the beginning of GI2 and drops from the end of GI2 so that the center of the plateau shall correspond to the middle point of GI2. While the work in 
\cite{bloessl2018performance} use the cross-correlation method for a SISO packet, we have preferred the auto-correlation method for the MIMO packet. 
 
Based on the fine-timing result, 
assume that $\hat n_2$ is the starting point of the first LTF symbol. A refined estimate of 
CFO in $\tilde r[n]$ is estimated as:
\begin{equation}
\hat\epsilon_2 =  \frac{ \text{phase}\Big\{\displaystyle\sum_{l=0}^{N_{\rm LTF} - 1} 
(\tilde r[\hat n_2 {+}l])^*  
\tilde r[\hat n_2 {+} l {+} N_{\rm LTF}]\Big\}
}{N_{\rm LTF}}.
\end{equation}

Next, the receiver performs channel estimation. For the data samples corresponding to 
the two LTF symbols, marked as 
LTF1 and LTF2, first refine the CFO compensation via
   $
    y[n]= e^{-j n \hat \epsilon_2 } \tilde r[n].
   $
   FFT is carried out on the two 
   LTF symbols to obtain 
   frequency-domain measurements:
 $ Y_{\rm LTF1}[k]$ 
 and $ Y_{\rm LTF2}[k]$ 
 on subcarrier $k$. 
 Let $ S_{\rm LTF}[k]$ denote the LTF symbol on subcarrier $k$ and $H_{\rm equ}[k]$ as the 
frequency response of 
$h_{\rm equ}(t)$.
We have:
\begin{equation}
    Y_{\rm LTF1}[k] = H_{\rm equ}[k] S_{\rm LTF}[k] + W_{\rm LTF1}[k],
\end{equation}
\begin{equation}
    Y_{\rm LTF2}[k] = H_{\rm equ}[k] S_{\rm LTF}[k] + W_{\rm LTF2}[k],
\end{equation}
which leads to 
a channel estimate as 
\begin{equation}
    \hat H_{\rm equ}[k] =  \frac{Y_{\rm LTF1}[k] + Y_{\rm LTF2}[k]}{2 
    S_{\rm LTF}[k]}.
    \label{eq:chest-LTF}
\end{equation}
Note that various 
improved channel estimation methods 
are available that improves
the accuracy exploiting the channel 
correlation across subcarriers  \cite{WIME}. 
These variations can be left for future development.

The data samples following  
L-LTF will be phase-compensated via:
\begin{equation}
    y[n]= e^{-j n \hat \epsilon_{\rm overall} } r[n],
\end{equation}
where the overall CFO is 
\begin{equation} 
\hat\epsilon_{\rm overall} = \hat\epsilon_1 + \hat\epsilon_2. 
\end{equation}

\medskip

\subsubsection{Determining Packet Formats}

The receiver next demodulates and decodes the next 
three OFDM symbols. 
Note that residual phase variations across OFDM blocks are compensated using the four pilot tones. 
Specifically, the receiver computes the average phase difference of received pilots to the transmitted pilots and compensate all the sub carriers with the average phase to get the QAM constellations for following demodulation and decoding. For each received OFDM symbol, the input-output model is
\begin{equation}
    Y[n;k] = H_{\rm equ}[k] S[n;k] 
    +W[n;k].
\end{equation}
Channel equalization is performed via 
\begin{equation}
\hat S[n;k] = 
\frac{Y[n;k]}{\hat H_{\rm equ}[k]}.
\label{eq:siso}
\end{equation}

The first block following L-LTF is L-SIG
using BPSK modulation. After BPSK demodulating and hard or soft decoding, the MCS and length of the transmitted packet are recovered if the parity check passes.  If the decoded MCS value is larger than 0, the packet format is 
legacy, and the receiver proceeds data transmission with the corresponding MCS mode. 

The decoded MCS value 0 indicates the potential of MIMO transmission. 
The receiver decodes the next two OFDM symbols with either BPSK or QBPSK modulations.
Correct decoding can be declared via the 
embedded CRC8 bits.
The receiver differentiates three different 
scenarios: (1) HT packet if QPBSK is used on both two symbols, (2) VHT packet if BPSK is on the first symbol and QPBSK is on the second symbol, and (3) otherwise, legacy packet with BPSK modulation.

\subsection{SU-MIMO}
Once the HT packet format is decided, 
$N_{\rm ss}$ is read from HT-SIG1 and the number of LTF symbols is also decided. Parallel data streams are transmitted which also necessitates the signals at multiple antennas at the receiver to be used. For a 2\texttimes2 MIMO, $N_{\rm TX} = 2$ at the AP and $N_{\rm RX} = 2$ at a station. 

For the $j$th receive antenna, the received signal can be expressed as 
\begin{equation}
     y_j(t) = 
     \sum_{i=1}^{N_{\rm TX}} x_i(t) \star h_i(t) + w_j(t)
 \end{equation}
 after CFO compensation. 
 With FFT processing on the $n$th OFDM block,
 the discrete-time model is
\begin{equation}
\underbrace{
    \begin{bmatrix}
        Y_1[n;k] \\
        Y_2[n;k] 
    \end{bmatrix} 
    }_{:=\bbY[n;k]}
    = 
    \underbrace{
    \begin{bmatrix}
        H_{11}[k] & H_{12}[k] \\
        H_{21}[k] & H_{22}[k] \\
    \end{bmatrix} 
    }_{:=\bbH[k]}
    \underbrace{
    \begin{bmatrix}
        S_1[n;k] \\
        S_2[n;k] 
    \end{bmatrix} 
    }_{:=\bbS[n;k]}
    {+} 
    \underbrace{\begin{bmatrix}
        W_1[n;k] \\
        W_2[n;k] \\
    \end{bmatrix}
    }_{:=\bbW[n;k]}
    \label{eq:SU-MIMO}
\end{equation}
MIMO channel estimation and data detection is carried out based on the input-output model in \eqref{eq:SU-MIMO}.

\medskip
\subsubsection{Processing HT-LTF}
For the HT-LTFs, each symbol $S_{\rm HT\text{-}LTF}[k]$ is assigned a specific polarity so that the transmitted HT-LTF symbols 
for each subcarrier $k$ as $S_{\rm HT\text{-}LTF}[k] \bbP$, where the  
orthogonal mapping matrix $\bbP$ in the 2\texttimes2 system is 
 \begin{equation}
   \bbP= 
    \begin{bmatrix}
        1 & -1\\
        1 & 1\\
    \end{bmatrix}. 
\end{equation}
Let $Y_{i,\text{LTF}j[k]}$ is the received signal at the $i$-th receive antenna for $j$-th LTF symbol 
on subcarrier $k$. Collecting the data from two HT-LTF symbols, one obtains: 
\begin{multline}
\underbrace{
    \begin{bmatrix}
        Y_{1,\text{LTF1}}[k] &
        Y_{1,\text{LTF2}}[k] \\
        Y_{2,\text{LTF1}}[k] &
        Y_{2,\text{LTF2}}[k] 
    \end{bmatrix} }_{:= \bbZ[k]}\\
    = 
\bbH[k]
\bbP S_{\rm HT\text{-}LTF}[k]
    + \begin{bmatrix}
        W_{1,\text{LTF1}}[k] &
        W_{1,\text{LTF2}}[k] \\
        W_{2,\text{LTF1}}[k] &
        W_{2,\text{LTF2}}[k] 
    \end{bmatrix}.
\end{multline}
The MIMO channel on subcarrier $k$ is estimated as 
\begin{equation}
    \hat \bbH[k] = 
    \frac{1}{S_{\rm HT\text{-}LTF}[k]}
    \bbZ[k]  \bbP^{-1}.
\end{equation}

\medskip
\subsubsection{Processing HT Data Blocks}
On each subcarrier, linear MIMO detection is carried out as
\begin{equation}
    \hat \bbS[n;k] = \left(\hat \bbH^H[k] \hat \bbH[k] \right)^{-1} \hat \bbH^H[k] \bbY[n;k].
\end{equation}
On each demodulated symbols $\hat S_i[n;k]$,
we adopt the simplified demapper of 16-QAM and 64-QAM constellations~\cite{tosato2002simplified} to 
generate the Log-Likelihood Ratio (LLR) of each coded bit. The LLRs are fed into a Viterbi decoder with soft decision, in contrast to the Viterbi decoder with hard decision adopted~\cite{WIME}.

\subsection {MU-MIMO}

The 2\texttimes2 MU-MIMO implementation is specified to one AP with two antennas and two stations with a single antenna, which are $N_{\rm TX}=2$ 
at the AP and $N_{\rm RX}=1$ per station. 
In the following section, we will present the two-way sounding and packet receiving procedure in detail. 

\medskip
\subsubsection{Channel Sounding via NDP} 

MU-MIMO in our implementation uses the VHT packet format. The NDP packet includes no data symbols and serves only for channel estimation. When broadcasting the NDP from AP to stations, no precoding is applied. Receivers measure the channel using two symbol of VHT LTFs. 
The estimated channel from STA1 is: 
\begin{equation}
    \begin{bmatrix}
    \hat H_{1}^{(1)}[k] & \hat H_{2}^{(1)}[k] 
    \end{bmatrix} 
    = 
    \frac{    \begin{bmatrix}
     Y_{\text{VHT-LTF1}}^{(1)}[k] & Y_{\text{VHT-LTF2}}^{(1)}[k] 
    \end{bmatrix}
    \bbP^{-1} }{S_{\text{VHT-LTF}}[k]}.
\end{equation} 
The estimated channel from STA2 is: 
\begin{equation}
    \begin{bmatrix}
    \hat H_{1}^{(2)}[k] & \hat H_{2}^{(2)}[k] 
    \end{bmatrix} 
    = 
    \frac{    \begin{bmatrix}
     Y_{\text{VHT-LTF1}}^{(2)}[k] & Y_{\text{VHT-LTF2}}^{(2)}[k] 
    \end{bmatrix}
    \bbP^{-1}}{S_{\text{VHT-LTF}}[k]}.
\end{equation}
STA1 and STA2 send back the 
estimated channels back to the access point as illustrated in Fig.~\ref{fig:diagram}.
While the standard defines multiple ways of channel feedback for VHT MU-MIMO, 
our current implementation sends back 
uncompressed CSI data.
    
\medskip
\subsubsection{VHT Precoded Data Transmission} 

Based on the channel feedback from STA1 and STA2,
the beamformer prepares beamforming vectors for data streams targeted to different users. For data stream to user 1, the beamforming vector is

\begin{equation}
\begin{split}
   \begin{bmatrix}
        Q_{11}[k] \\
        Q_{21}[k] 
    \end{bmatrix} 
    &=
        \frac{1}{\sqrt{\big|\hat H^{(2)}_{2}[k]\big|^2 + \big|\hat H^{(2)}_{1}[k]\big|^2}} 
    \begin{bmatrix}
        - \hat H_{2}^{(2)}[k] \\
        \ \hat H_{1}^{(2)}[k]  
    \end{bmatrix} 
\end{split}
\end{equation}
which is orthogonal to the channel of STA2:
$ [\hat H_{1}^{(2)}[k],   \hat H_{2}^{(2)}[k] ]^T $.
For the data stream to STA2, the beamforming vector is
\begin{equation}
\begin{split}
   \begin{bmatrix}
         Q_{12}[k] \\
         Q_{22}[k] 
    \end{bmatrix} 
    &=
 \frac{1}{\sqrt{\big|\hat H^{(1)}_{2}[k]\big|^2 + \big|\hat H^{(1)}_{1}[k]\big|^2}}  
    \begin{bmatrix}
         - \hat H_{2}^{(1)}[k] \\
         \hat H_{1}^{(1)}[k] 
    \end{bmatrix} 
\end{split}
\end{equation}
which is orthogonal to the channel of STA1:
$ [\hat H_{1}^{(1)}[k],   \hat H_{2}^{(1)}[k] ]^T$.

With precoding, the received signal at STA1 is 
\begin{multline}
   Y^{(1)}[n;k] = 
   \underbrace{(H_{1}^{(1)}[k]Q_{11}[k]
   +H_{2}^{(1)}[k]Q_{21}[k])}_{H_{\text{VHT}}^{(1)}[k]} S^{(1)}[n;k]\\
   + Z^{(1)}[n;k]
   \label{eq:sta1}
\end{multline}
where $Z^{(1)}[n;k]$ is the equivalent noise for STA1.
Similary for STA2, we have 
\begin{multline}
   Y^{(2)}[n;k] = 
   \underbrace{(H_{1}^{(2)}[k]Q_{12}[k]
   +H_{2}^{(2)}[k]Q_{22}[k])}_{H_{\text{VHT}[k]}^{(2)}}
   S^{(2)}[n;k] \\
   + Z^{(2)}[n;k]. 
   \label{eq:sta2}
\end{multline}

The precoding is applied on VHT-STF, VHT-LTF, and VHT data symbols.
\textit{For these symbols, 
the receiver 
estimates the equivalent
channel $H_{\text{VHT}}^{(i)}[k]$
in \eqref{eq:sta1} and \eqref{eq:sta2},
and channel equalization is applied 
based on the estimated equivalent channel.}
The receiver procedure for each station is the same as 
if this is a single data stream transmission.
Specifically, channel estimation is carried using
expression in \eqref{eq:chest-LTF}
and data demodulation is carried out using
expression in \eqref{eq:siso}, 
where the interference from the other user
is treated as additive noise.

\section{GR-WiFi Implementation}\label{sec:implement}

\begin{figure*}[ht]
    \centering
    \includegraphics[width=1\linewidth]{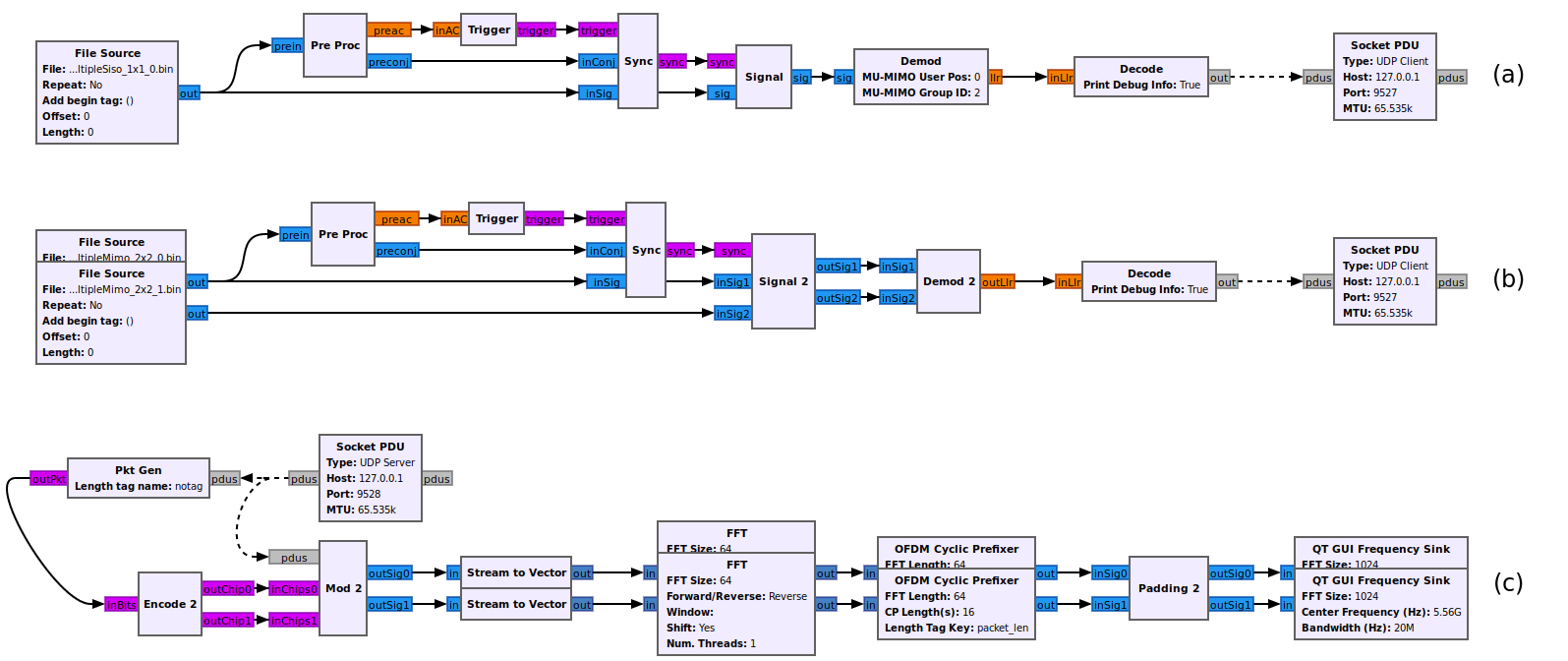}
    \caption{GNU Radio implementations of (a) SISO receiver, (b) 2\texttimes2 MIMO receiver and (c) 2\texttimes2 MIMO transmitter.}
    \label{fig:grimpl}
\end{figure*}

\begin{figure*}[ht]
    \centering
    \includegraphics[width=1\linewidth]{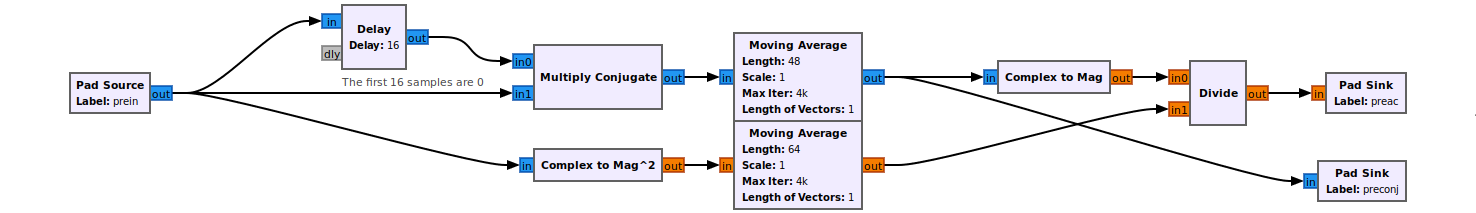}
    \caption{Components of the pre-processing block.}
    \label{fig:grpreproc}
\end{figure*}

In this section, we introduce the implementation details of the GR-WiFi system. The system includes the full physical layer implemented on GNU Radio with transmitter and receivers supporting up to 2\texttimes2 SU-MIMO and 2\texttimes2 MU-MIMO (AP transmitting 2 spatial streams to 2 stations) and a simple media access layer (MAC) implemented by Python. To make an easier maintenance and future development and extensions, the MU-MIMO channel sounding and beamforming spatial mapping matrix generation is also implemented by Python.

Fig.~\ref{fig:grimpl} shows the implemented SISO and 2\texttimes2 MIMO receivers. The 2\texttimes2 MIMO is compatible with SISO packet reception. However, we still make an individual SISO block because the receiver keeps processing samples all the time. If the MIMO receiver is applied to a SISO platform, the samples on the second port are useless and waste computing resources. For the transmitter, it only generates the samples of packets, so the wasted resource is negligible when it is applied to an SISO transmitter.

In the implementation, all the blocks are developed using C++ and the samples are processed in serial. In the following, we describe the detail of each block.

\subsubsection{Pre-Processing}
The pre-processing block is a hierarchical block and its details are presented in Fig.~\ref{fig:grpreproc}. The main function of this block is to compute the auto-correlation of the input samples. And the values during the auto-correlation computation is reused to compute the coarse CFO. The average blocks use sliding window with maximum length to avoid repeated computation of samples while preventing accumulated error of float numbers.

\subsubsection{Trigger}
The trigger block takes the continuous auto-correlation samples as the input and detects the plateau of auto-correlation. Once the length of plateau is long enough, it generates a flag in the output to trigger the synchronization block. The trigger block also has a state machine to avoid multiple triggers for one packet to save computing resources.

\subsubsection{Synchronization}
The synchronization block is triggered to use auto-correlation of LTF and find the beginning of LTF whose index is passed to the next block. At the same time, this block compensates the coarse CFO from pre-processing block to the LTF samples, re-estimates the CFO with two LTF symbols and computes the accurate CFO value. However, the CFO for the packet samples is not compensated here since the packet length is not clear and if all the following samples are compensated the computing resources could be wasted. The CFO value is passed to the next block using a tag.

\subsubsection{Signal}
The signal block is triggered by the synchronization block to get the timing of the packet and the estimated CFO. It first compensates the CFO for LTF and Legacy Signal to estimate the Legacy channel and demodulate and decode the Legacy Signal. If the Legacy Signal is correct, this packet is identified to be at least a Legacy packet with the maximum possible timing length corresponding to the Legacy length. The signal block computes the symbol number and sample number according to the MCS and packet length. The following input samples within the sample number will be compensated with CFO and output to the next block. That means the signal block chops the sample stream and only keeps the useful packet samples to the following processing. The Legacy channel is also passed to the next block with a tag.

For the MIMO receivers, the Signal2 block is applied while the only difference is that the Signal2 block also chops samples from the 2nd sample stream with the same length as the 1st stream and the CFO of the 2nd stream is also compensated.

\subsubsection{Demodulation}
The demodulation block demodulates the OFDM symbols to QAM constellations and disassembles them into soft bits. It first receives the Legacy channel and then checks the MCS. If the MCS is the lowest, it further demodulates and decodes another two symbols to check the HT Signal and VHT Signal A which will decide the following demodulation. If the Non-Legacy checking fails, the packet is demodulated as Legacy. If it is either HT or VHT, the channel is re-estimated and then the packet is demodulated. For the OFDM part, it compensates the channel after FFT and uses pilots to compensate the residual CFO. The QAM constellations are disassembled into soft bits and deinterleaved. To speed up the processing, some steps are simplified such as the deinterleaving is using a lookup table for each symbol but not following the method given in standards. The demodulation block outputs the soft bits to the next decoding block.

The difference between the demodulation block and demodulation2 block is that the demodulation is designed for SISO and also the receiver of MU-MIMO. It has the ability to sound channel and receive the MU-MIMO packet with group number and position in the group to estimate the corresponding channel and demodulate accordingly. For the demodulation2 block designed for the AP side in MU-MIMO, we remove the channel sounding and MU-MIMO reception to simplify the processing. However, we also consider adding full functionality to all the blocks in future development.

\subsubsection{Decoding}
The decoding block only has an input stream port and the output port is message type but not stream type. It takes the input soft bits with given trellis length, decodes the packet and checks the CRC32. The Viterbi decoder updates forward for the whole packet and then traces back from the very end which is the 6-bit-zero tail.  
If the packet is correct then it reports the packet to the MAC layer implemented using Python through a UDP message for further customized processing. For an NDP packet, which contains no data symbols, it packs the the channel state information into UDP message and sends it to the MAC layer. This enables the MAC layer to perform channel sounding process.

\subsubsection{Compatibility and Extensibility}
The proposed receiver structure not only works for basic OFDM symbol length of Legacy, HT and VHT physical layer formats in IEEE 802.11a/g/n/ac, it also works for short guard interval in HT and VHT. Besides, this structure can also be extended to support IEEE 802.11ax/be (WiFi 6 and 7) which have the similar Legacy compatible format in the preamble. The compatibility and extensibility enable the vitality of this platform to support future evaluation and research work.

\section{Performance Evaluation }\label{sec:performance}
In this section, we present the performance evaluation of GR-WiFi through simulations and real-world testbed experiments. 

\subsection{SISO}
\subsubsection{Simulation  Studies}
We first compare the performance of GR-WiFi on Legacy transmissions with WIME~\cite{bloessl2018performance}. To compute the PDR result on each SNR value, 1000 UDP packets with a random 500-byte payload for each packet are generated and white Gaussian noise with the amplitude corresponding to the given SNR is added to the signal. We input the same generated signal to both GR-WiFi and WIME.

\begin{figure}[t]
\centering
\includegraphics[width=0.99\linewidth]{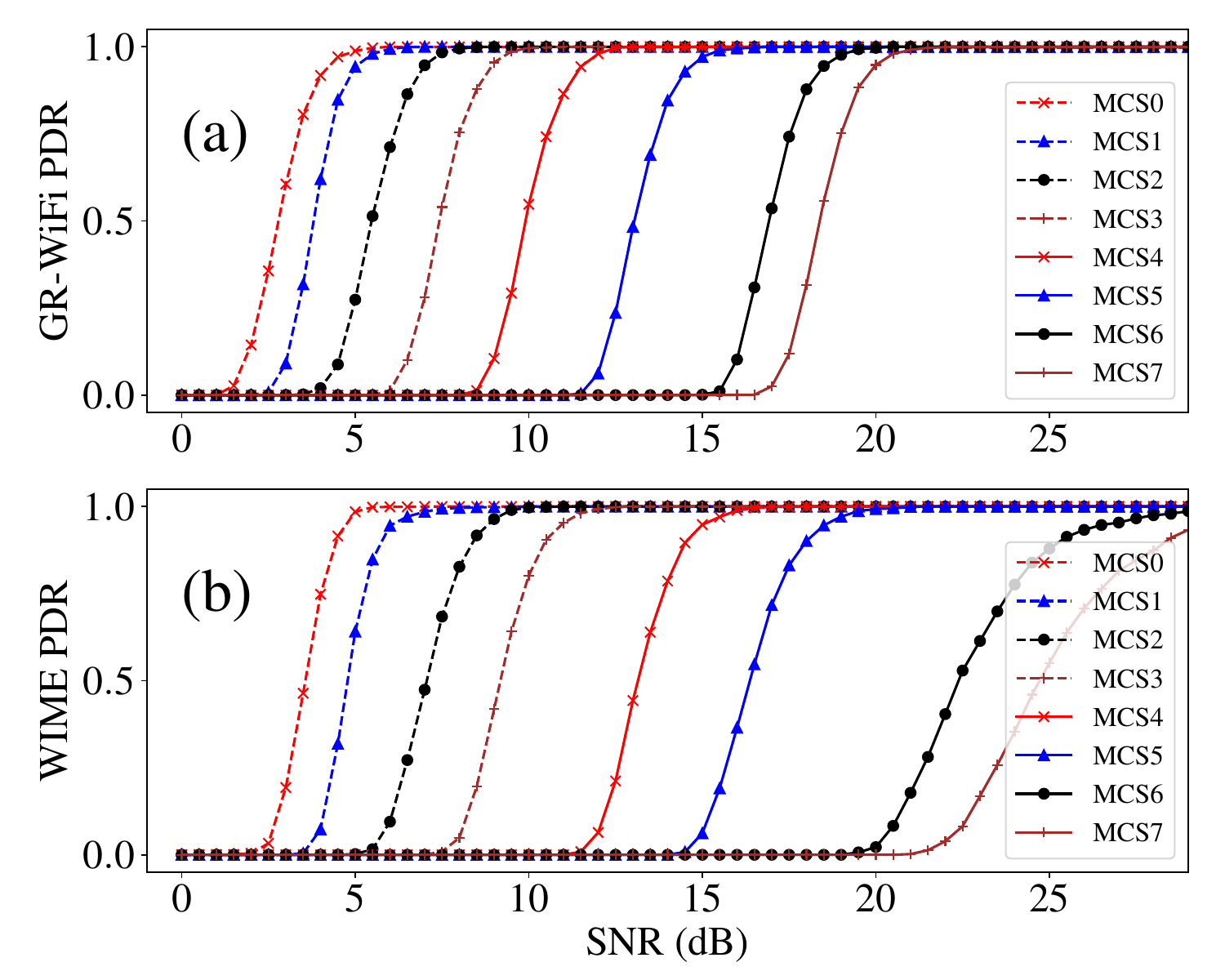}
\caption{Simulation performance on PDR against SNR for the comparison between GR-WiFi and WIME using Legacy format.}
\label{fig:perflcompare}
\end{figure}

As shown in Fig.~\ref{fig:perflcompare}, compared to WIME, the proposed GR-WiFi has a better gain from 1 dB to 3.5 dB for MCS 0-5, and a more stable performance for MCS 6-7. Next, we compare the performance on CFO tolerance of the two receivers. We add the CFO of 233 kHz for signal generation, which is the tolerance as required by the standard. 
\begin{figure}[t]
\centering
\includegraphics[width=0.99\linewidth]{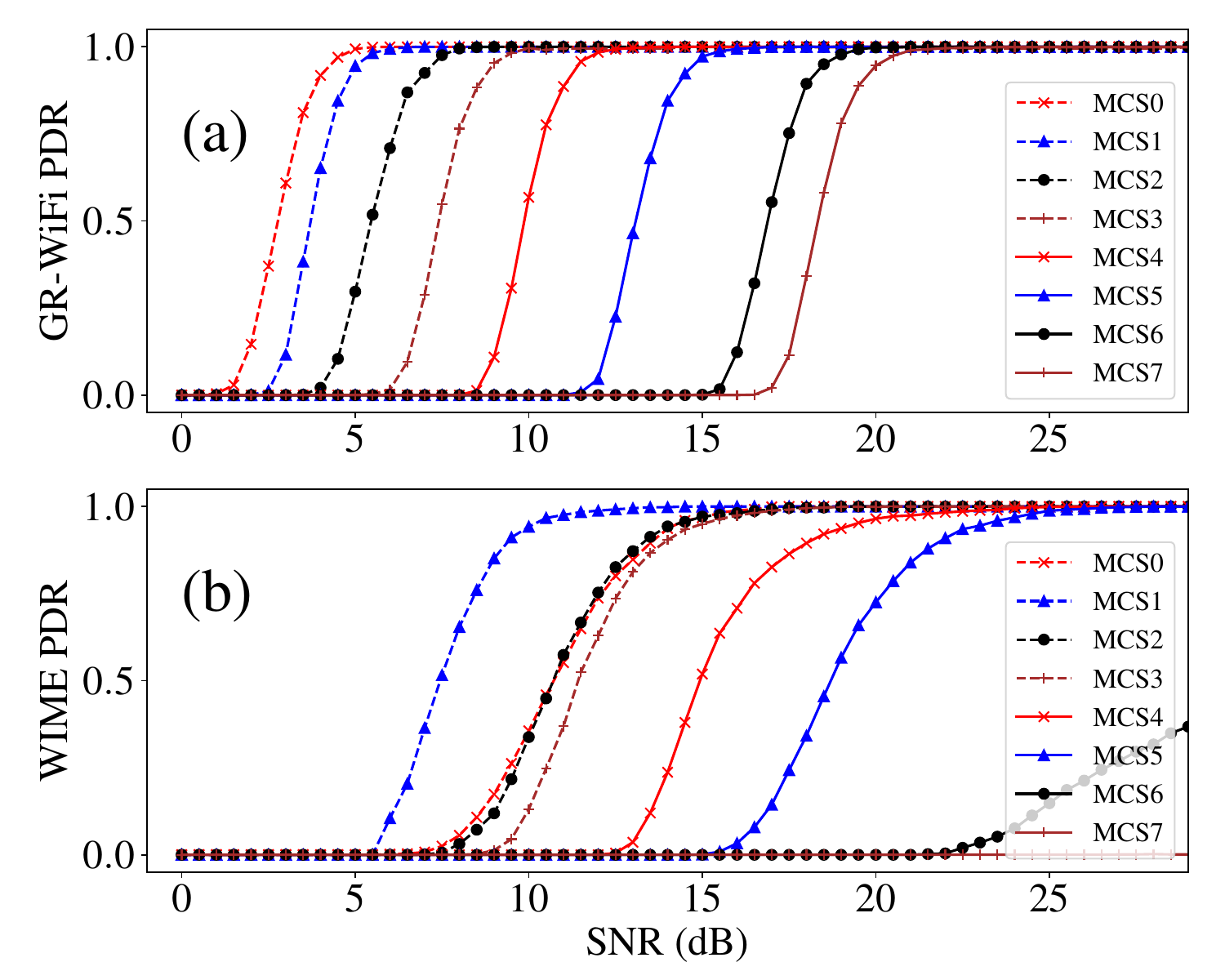}
\caption{Simulation performance on PDR against SNR for the comparison between GR-WiFi and WIME using Legacy format with 233 kHz CFO.}
\label{fig:perflcomparecfo}
\end{figure}

As shown in Fig.~\ref{fig:perflcomparecfo}, the proposed GR-WiFi retains similar performance under the condition with CFO, while performance of WIME is not stable under such condition.

\begin{figure}[t]
\centering
\includegraphics[width=0.99\linewidth]{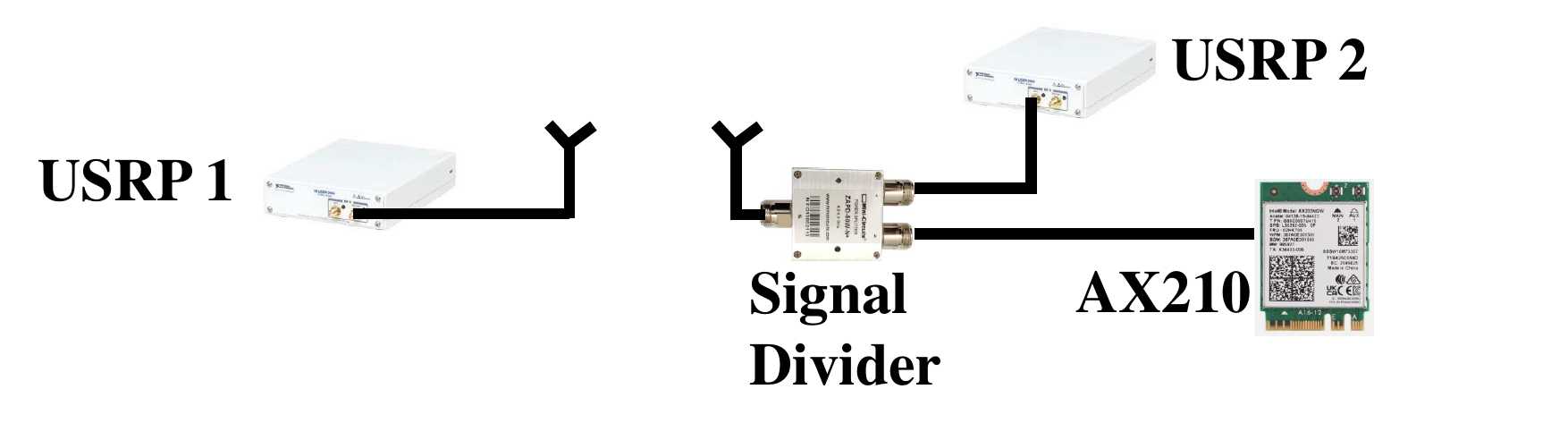}
\caption{Testbed setup for the VHT SISO performance comparison between GR-WiFi and AX210.}
\label{fig:perfcommercialtestbed}
\end{figure}

\subsubsection{Comparison 
with Commercial Adapter}

\begin{figure}[t]
\centering
\includegraphics[width=0.99\linewidth]{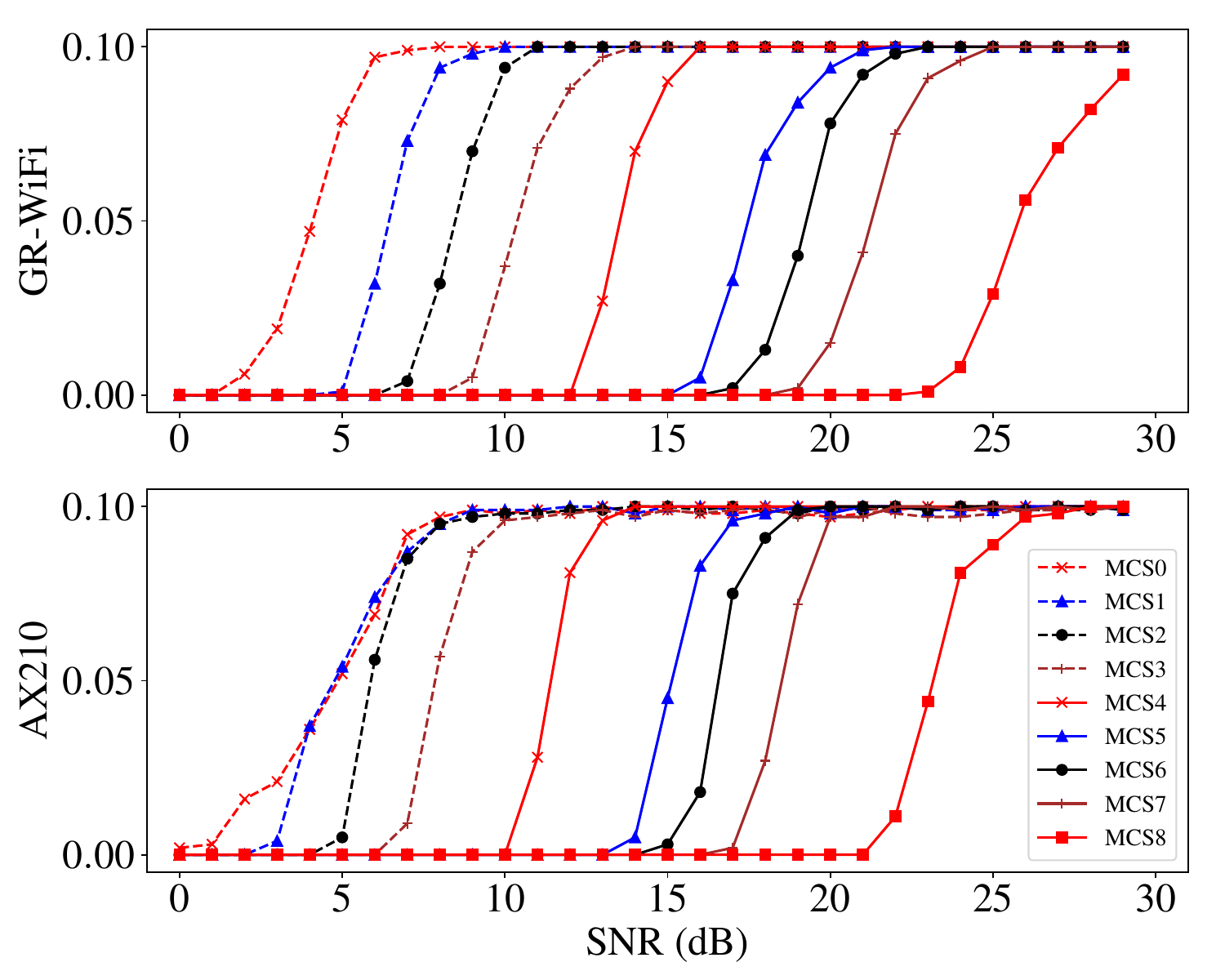}
\caption{Field test performance on the comparison of VHT SISO between GR-WiFi and AX210. The data rates for MCS0-8 in 802.11a/g are: 6, 9, 12, 18, 24, 36, 48, and 54 Mb/s}
\label{fig:perfcommercial}
\end{figure}

To show the performance of GR-WiFi in real field, we compare its performance with Intel AX210 wireless adaptor. In this experiment, we use one USRP 1 to send VHT packets with added Gaussian noise to generate packets with different SNR on one side. On the other side, we use a signal divider to receive the signal and output two same signals to both the AX210 and USRP 2 running GR-WiFi. The testbed in shown in Fig.~\ref{fig:perfcommercialtestbed}, and the results are presented in Fig.~\ref{fig:perfcommercial}. From the results, we can observe that AX210 performs worse than GR-WiFi for MCS 0 and 1. However, for the other MCS, the AX210 is around 2 dB better than GR-WiFi. This result shows that the sensitivity of GR-WiFi is close to commercial devices.

\begin{figure}[t]
\centering
\includegraphics[width=0.99\linewidth]{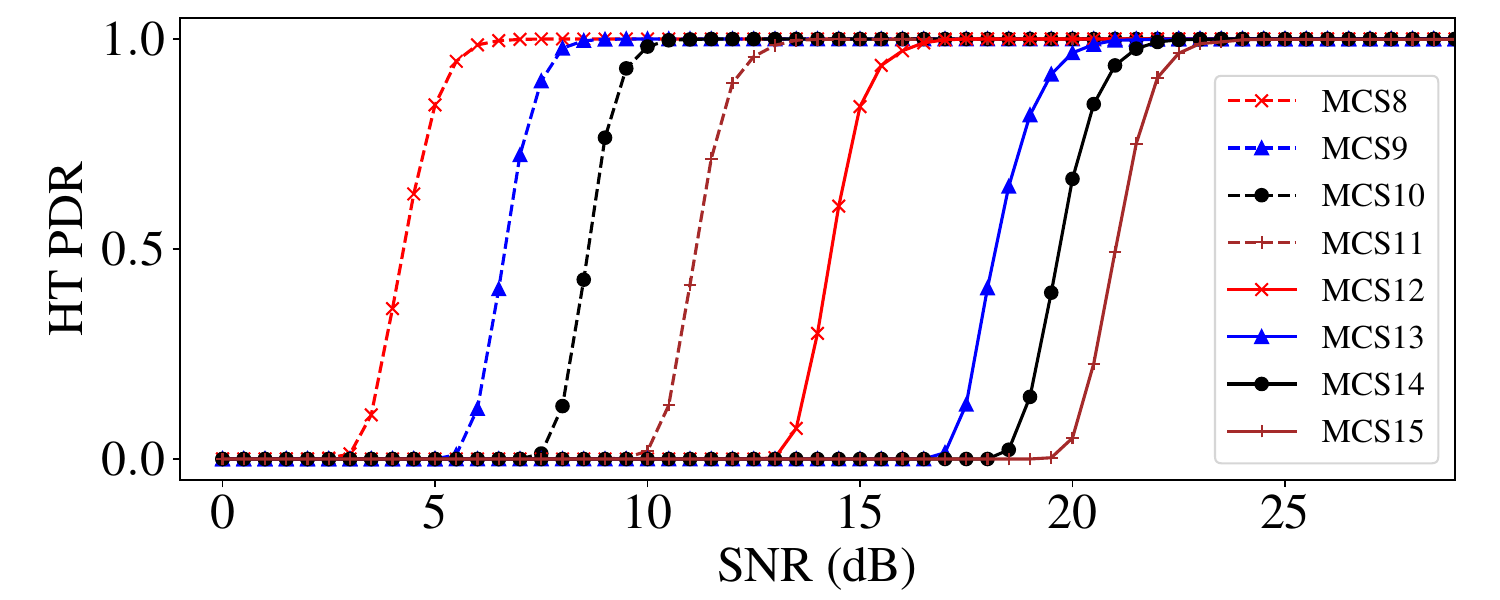}
\caption{Simulation performance on SU-MIMO HT PDR against SNR of modulation coding schemes with AWGN, CFO = 0 Hz. The data rates for MCS8-15 in 802.11n are: 13, 26, 39, 52, 78, 104, 117, and 130 Mb/s}
\label{fig:perfnlsumimo}
\end{figure}

\begin{figure}[t]
\centering
\includegraphics[width=0.99\linewidth]{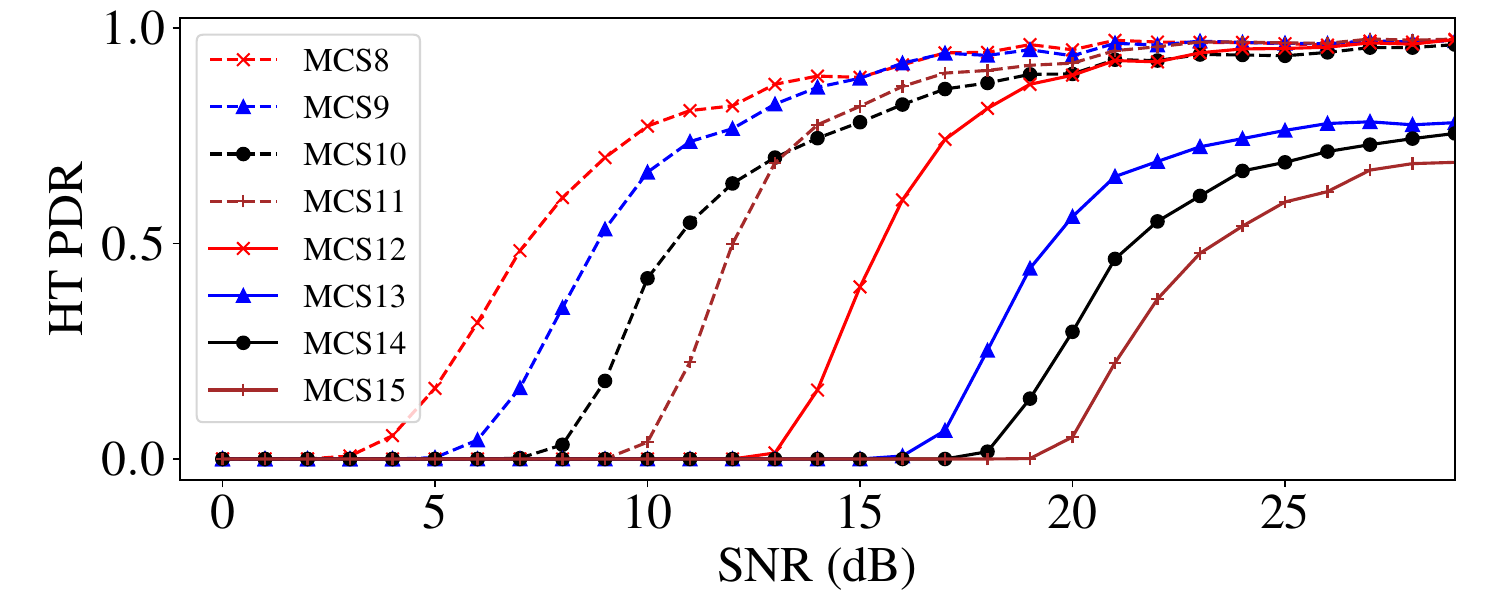}
\caption{Simulation performance on SU-MIMO HT PDR against SNR of modulation coding schemes with AWGN and TGac Channel Model B appiled.}
\label{fig:perfnlsumimochanB}
\end{figure}

\subsection{SU-MIMO}
Fig.~\ref{fig:perfnlsumimo} shows the simulation performance of the HT SU-MIMO transmissions with AWGN noise and no CFO. For SU-MIMO with two spatial streams and EQM rates, the MCS has value from 8 to 15. Comparing the SU-MIMO performance 
to that of SISO, we observe that higher SNR is required to achieve same PDR. This is due to the inter-stream interference from the simultaneous transmission of multiple spatial streams. However, under high SNR conditions, MIMO transmissions provide higher data rates.  

Fig.~\ref{fig:perfnlsumimochanB} shows the simulation performance for HT SU-MIMO in the presence of AWGN and the TGac Channel Model B. Unlike Fig.~\ref{fig:perfnlsumimo}, which simulates an ideal AWGN channel, this model introduces multipath fading and frequency selectivity, reflecting the complexities of real-world indoor environments. As the result, the system experiences performance degradation. It requires higher SNR to achieve comparable packet delivery rates. The degradation is particularly significant at higher MCS levels, emphasizing the performance gap between ideal conditions and realistic simulated scenarios.

\subsection{MU-MIMO}

\begin{figure}[t]
\centering
\includegraphics[width=0.99\linewidth]{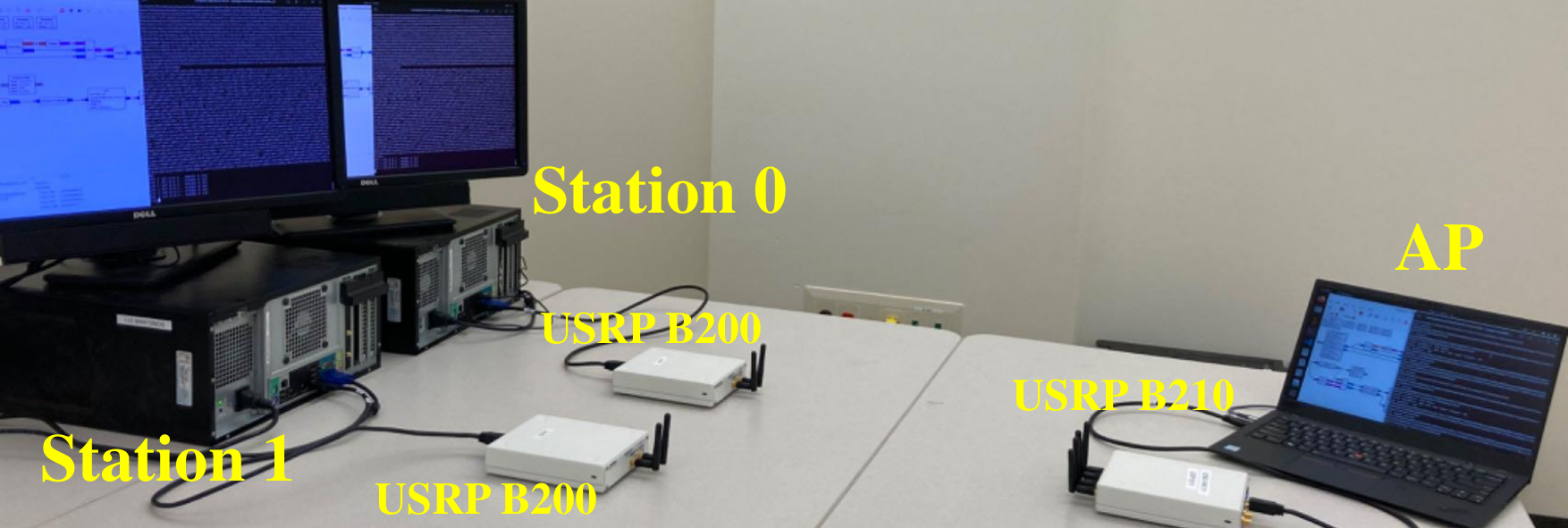}
\caption{MU-MIMO testbed with one AP and two stations.}
\label{fig:mumimotestbed}
\end{figure}

The performance of MU-MIMO is evaluated in a field test. Fig.~\ref{fig:mumimotestbed} shows the testbed setup in our lab including one USRP B210 with 2 TX and 2 RX as the AP, and two USRP B200 with 1 TX and 1 RX as the stations. During the experiments, all the devices are stationary. 

\begin{figure}[t]
\centering
\includegraphics[width=0.99\linewidth]{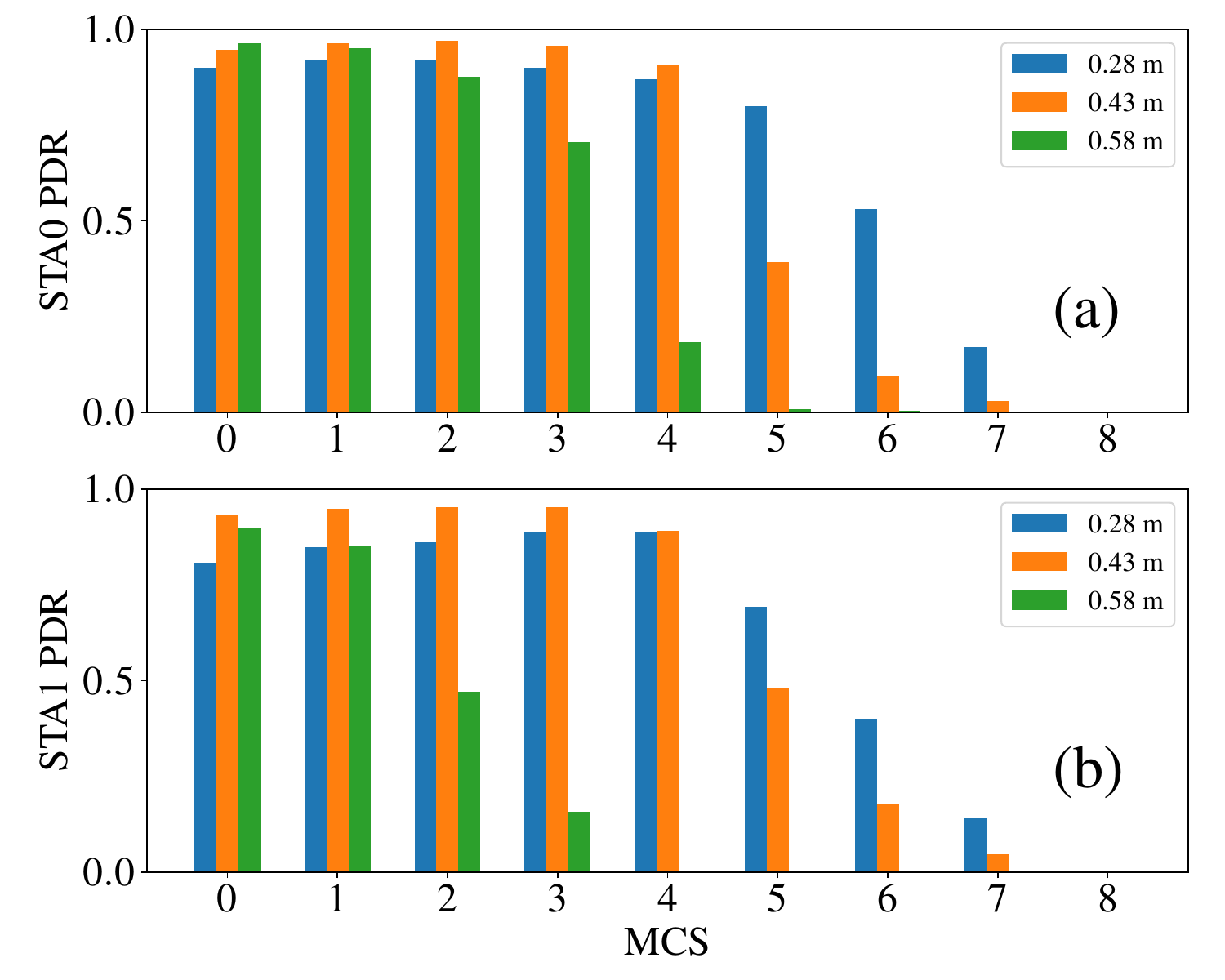}
\caption{Field test performance on PDR against MCS of MU-MIMO.}
\label{fig:perfmufield}
\end{figure}

Fig.~\ref{fig:perfmufield} shows the results of the MU-MIMO experiments with all MCS tested. 
For each MCS that has successful channel sounding, 100 packets with a 1 $m$s gap between each two packets are transmitted. At each distance, This process is repeated 100 times, and the average PDR is computed for each MCS. 
The performance at three different AP-to-station distances for both stations are showed in Fig.~\ref{fig:perfmufield}(a) and (b). Based on the results, we observe that as the distance increases from 0.28 m to 0.58 m, the PDR tends to decrease across the MCS values, especially at higher MCS. At MCS 0, 1, and 2, the PDR remains close to 1.0 for all three distances in both STA0 and STA1, indicating stable delivery at lower MCS. As the MCS index increases, the PDR starts to decrease. At MCS 5, 6, and 7, the PDR drops sharply, with longer distances (0.43 m and 0.58 m) showing almost no successful delivery. MCS 8 has zero PDR across all distances since its high modulation scheme (256-QAM) requires a very high SNR to achieve successful packet delivery. 

\section{Consclusion and Future work}\label{sec:conclusion}

In this paper, we presented an open-source GNU Radio transceiver, GR-WiFi, for the IEEE 802.11n/ac physical layer supporting 2\texttimes2 SU-MIMO and MU-MIMO. We presented the principle of IEEE 802.11 OFDM transmissions, receptions and channel sounding, and compared the performance of the proposed receiver with not only state-of-the-art solutions but also COTS wireless adaptors. The results show that the proposed transceiver achieves good performance and also fully supports the coexistence of 
legacy, HT and VHT traffics defined in IEEE 802.11n/ac standards.

The open-source GNU Radio package provides foundation for us and the community users to pursue various future research topics. Examples include (a) 
support more than 2 spatial streams 
for high data rates; (b) evaluation of the impact of feedback delay
and feedback quantization on MU-MIMO systems; (c) network protocol design leveraging MU-MIMO physical layer to support low-latency traffics; (d) support of newer features, e.g., orthogonal frequency-division multiple access (OFDMA) in IEEE 802.11ax, and (e) development of novel multiple access schemes, e.g., rate splitting multiple access (RSMA)~\cite{XL_BC_2024_RSMAProto},
for NextG wireless systems.

\section*{ACKNOWLEDGEMENT}
This work is supported by NSF Awards CNS-1932480, CNS-2008463, CCF-2028875, ECCS 2332172 and UConn SPARK award.  We are grateful to Dr. Bloessl whose open-source GNU-radio receiver in \cite{WIME} has motivated this work.

\vspace*{4\baselineskip}




\ifCLASSOPTIONcaptionsoff
  \newpage
\fi


%
\bibliographystyle{IEEEtran}
\bibliography{references,bibliography}

\begin{thebibliography}{10}
\providecommand{\url}[1]{#1}
\csname url@samestyle\endcsname
\providecommand{\newblock}{\relax}
\providecommand{\bibinfo}[2]{#2}
\providecommand{\BIBentrySTDinterwordspacing}{\spaceskip=0pt\relax}
\providecommand{\BIBentryALTinterwordstretchfactor}{4}
\providecommand{\BIBentryALTinterwordspacing}{\spaceskip=\fontdimen2\font plus
\BIBentryALTinterwordstretchfactor\fontdimen3\font minus \fontdimen4\font\relax}
\providecommand{\BIBforeignlanguage}[2]{{%
\expandafter\ifx\csname l@#1\endcsname\relax
\typeout{** WARNING: IEEEtran.bst: No hyphenation pattern has been}%
\typeout{** loaded for the language `#1'. Using the pattern for}%
\typeout{** the default language instead.}%
\else
\language=\csname l@#1\endcsname
\fi
#2}}
\providecommand{\BIBdecl}{\relax}
\BIBdecl

\bibitem{8024171}
L.~{Davoli}, L.~{Belli}, A.~{Cilfone}, and G.~{Ferrari}, ``From {Micro} to {Macro} {IoT:} {Challenges} and solutions in the integration of {IEEE} 802.15.4/802.11 and {Sub-GHz} technologies,'' \emph{IEEE Internet of Things Journal}, vol.~5, no.~2, pp. 784--793, 2018.

\bibitem{sisinni2018industrial}
E.~Sisinni, A.~Saifullah, S.~Han, U.~Jennehag, and M.~Gidlund, ``Industrial internet of things: Challenges, opportunities, and directions,'' \emph{IEEE transactions on industrial informatics}, vol.~14, no.~11, pp. 4724--4734, 2018.

\bibitem{bloessl2013gnu}
B.~Bloessl, C.~Leitner, F.~Dressler, and C.~Sommer, ``{A GNU Radio-based IEEE 802.15.4 Testbed},'' in \emph{12. GI/ITG KuVS Fachgespr{\"{a}}ch Drahtlose Sensornetze (FGSN 2013)}, September 2013, pp. 37--40.

\bibitem{9348953}
E.~Faulkner, Z.~Yun, S.~Zhou, Z.~J. Shi, S.~Han, and G.~B. Giannakis, ``{An Advanced GNU Radio Receiver of IEEE 802.15.4 OQPSK Physical Layer},'' \emph{IEEE Internet of Things Journal}, vol.~8, no.~11, pp. 9206--9218, 2021.

\bibitem{gast2005802}
M.~Gast, \emph{{802.11 Wireless Networks: The Definitive Guide}}.\hskip 1em plus 0.5em minus 0.4em\relax O'Reilly Media, Inc., 2005.

\bibitem{9502043}
``{IEEE Standard for Information Technology--Telecommunications and Information Exchange between Systems - Local and Metropolitan Area Networks--Specific Requirements - Part 11: Wireless LAN Medium Access Control (MAC) and Physical Layer (PHY) Specifications},'' \emph{{IEEE Std 802.11-2020 (Revision of IEEE Std 802.11-2016)}}, pp. 1--7524, 2021.

\bibitem{SCHULZ2018269}
M.~Schulz, D.~Wegemer, and M.~Hollick, ``The nexmon firmware analysis and modification framework: Empowering researchers to enhance wi-fi devices,'' \emph{Computer Communications}, vol. 129, pp. 269--285, 2018.

\bibitem{bloessl2018performance}
B.~Bloessl, M.~Segata, C.~Sommer, and F.~Dressler, ``{Performance Assessment of IEEE 802.11p with an Open Source SDR-based Prototype},'' \emph{IEEE Transactions on Mobile Computing}, vol.~17, no.~5, pp. 1162--1175, May 2018.

\bibitem{grwifi}
\BIBentryALTinterwordspacing
``G{R}-{W}i{F}i.'' [Online]. Available: \url{https://github.com/cloud9477/gr-ieee80211}
\BIBentrySTDinterwordspacing

\bibitem{openwifi}
\BIBentryALTinterwordspacing
``Openwifi.'' [Online]. Available: \url{https://github.com/open-sdr/openwifi}
\BIBentrySTDinterwordspacing

\bibitem{nuandBladeRFwiphyNuand}
\BIBentryALTinterwordspacing
``Bladerf-wiphy.'' [Online]. Available: \url{https://www.nuand.com/bladerf-wiphy/"}
\BIBentrySTDinterwordspacing

\bibitem{WIME}
B.~Bloessl, ``{Wireless Measurement and Experimentation (WIME)},'' https://www.wime-project.net/.

\bibitem{9804669}
Z.~Yun, P.~Wu, S.~Zhou, A.~K. Mok, M.~Nixon, and S.~Han, ``{RT-WiFi on Software-Defined Radio: Design and Implementation},'' in \emph{Proc. of 28th IEEE Real-Time and Embedded Technology and Applications Symposium (RTAS)}, 2022, pp. 254--266.

\bibitem{4022536}
K.~Nishimori, R.~Kudo, Y.~Takatoti, A.~Ohta, and K.~Tsunekawa, ``{Performance Evaluation of 8×8 Multi-User MIMO-OFDM Testbed in an Actual Indoor Environment},'' in \emph{International Symposium on Personal, Indoor and Mobile Radio Communications}, 2006, pp. 1--5.

\bibitem{4489270}
T.~Wirth, V.~Jungnickel, A.~Forck, S.~Wahls, H.~Gaebler, T.~Haustein, J.~Eichinger, D.~Monge, E.~Schulz, C.~Juchems, F.~Luhn, and R.~Zavrtak, ``Real-time multi-user multi-antenna downlink measurements,'' in \emph{IEEE Wireless Comm. and Networking Conference}, 2008, pp. 1328--1333.

\bibitem{2010mobicom}
E.~Aryafar, N.~Anand, T.~Salonidis, and E.~W. Knightly, ``{Design and experimental evaluation of multi-user beamforming in wireless LANs},'' in \emph{Proceedings of the sixteenth annual international conference on Mobile computing and networking}, 2010, pp. 197--208.

\bibitem{8116431}
T.~Kuber, G.~Sridharan, D.~Saha, and I.~Seskar, ``{Practical MU-MIMO Experiments Using SDRs},'' in \emph{2017 IEEE Conference on Computer Communications Workshops (INFOCOM WKSHPS)}, 2017, pp. 517--522.

\bibitem{9259366}
P.~K. Sangdeh, H.~Pirayesh, A.~Mobiny, and H.~Zeng, ``{LB-SciFi: Online Learning-Based Channel Feedback for MU-MIMO in Wireless LANs},'' in \emph{the 28th IEEE International Conference on Network Protocols (ICNP)}, 2020, pp. 1--11.

\bibitem{10.1145/2973750.2973758}
S.~Sur, I.~Pefkianakis, X.~Zhang, and K.-H. Kim, ``{Practical MU-MIMO user selection on 802.11ac commodity networks},'' in \emph{Proceedings of the 22nd Annual International Conference on Mobile Computing and Networking (MobiCom)}, 2016, p. 122–134.

\bibitem{tosato2002simplified}
F.~Tosato and P.~Bisaglia, ``{Simplified Soft-Output Demapper for Binary Interleaved COFDM with Application to HIPERLAN/2},'' in \emph{IEEE International Conference on Communications}, vol.~2, 2002, pp. 664--668.

\bibitem{XL_BC_2024_RSMAProto}
X.~Lyu, S.~Aditya, J.~Kim, and B.~Clerckx, ``Rate-splitting multiple access: The first prototype and experimental validation of its superiority over {SDMA} and {NOMA},'' \emph{IEEE Transactions on Wireless Communications}, pp. 1--1, 2024.

\end{thebibliography}





\end{document}